\documentclass[
    amsmath,
    nofootinbib,
    amssymb,
    english,
    aps,prd,
    showkeys,
]{revtex4-2}
\usepackage{xcolor}
\usepackage{fix-cm}
\usepackage[utf8]{inputenc}
\usepackage[english]{babel}
\usepackage{graphicx}
\usepackage{latexsym,amsthm,amsfonts}
\usepackage{makeidx}   
\usepackage[T1]{fontenc}
\usepackage{subfigure}
\usepackage{dcolumn}
\usepackage{bm}
\usepackage{cancel}
\usepackage{xcolor}
\usepackage[hyperindex,colorlinks,linkcolor=blue,citecolor=red]{hyperref}

\usepackage{textgreek}
\usepackage{comment}

\allowdisplaybreaks[1]

\usepackage{upgreek}
\usepackage{graphicx}
\usepackage{tensor}

\begin{document}
\title{Sources of matter for wormholes in a k-essence theory}

 \author{Marcos V. de S. Silva\footnote{Author to whom any correspondence should be addressed.}}
	\email{marcos.sousa@uva.es}
\affiliation{Department of Theoretical Physics, Atomic and Optics, Campus Miguel Delibes, \\ University of Valladolid UVA, Paseo Bel\'en, 7,
47011 - Valladolid, Spain}
	\affiliation{Departamento de Física, Programa de Pós-Graduação em Física,
Universidade Federal do Ceará, Campus Pici, 60440-900, Fortaleza, Ceará, Brazil}	

\author{Carlos F. S. Pereira} 
	\email{carlosfisica32@gmail.com}
	\affiliation{Departamento de Física, Universidade Federal do Espírito Santo, Avenida Fernando Ferrari, 514, Goiabeiras, 29060-900, Vit\'oria, Espir\'ito Santo, Brazil.}

\author{Bruna Bragato}
	\email{bruna.bragato@edu.ufes.br}
	\affiliation{Departamento de Física, Universidade Federal do Espírito Santo, Avenida Fernando Ferrari, 514, Goiabeiras, 29060-900, Vit\'oria, Espir\'ito Santo, Brazil.}

\author{Manuel E. Rodrigues}
    \email[]{esialg@gmail.com}
	\affiliation{Faculdade de Ci\^encias Exatas e Tecnologia, Universidade Federal do Par\'a Campus Universit\'ario de Abaetetuba, 68440-000, Abaetetuba, Par\'a, Brazil and Faculdade de F\'isica, Programa de P\'os-Gradua\c{c}\~ao em F\'isica, Universidade Federal do Par\'a, 66075-110, Bel\'em, Par\'a, Brazil.}

\author{Júlio C. Fabris}
    \email[]{julio.fabris@cosmo-ufes.org}
	\affiliation{Núcleo Cosmo-ufes \& Departamento de Física, Universidade Federal do Espírito Santo, Avenida Fernando Ferrari, 514, Goiabeiras, 29060-900, Vit\'oria, Espir\'ito Santo, Brazil}
    \affiliation{National Research Nuclear University MEPhI (Moscow Engineering Physics Institute), 115409, Kashirskoe shosse 31, Moscow, Russia.}
	
	\author{H. Belich} 
    \email{humberto.belich@ufes.br}
    \affiliation{Departamento de F\'isica e Qu\'imica, Universidade Federal do Esp\'irito Santo, Avenida Fernando Ferrari, 514, Goiabeiras, Vit\'oria, Espir\'ito Santo, 29060-900, Brazil.}

\begin{abstract}  
In this work, we analyze some matter sources associated with wormhole models within a k-essence theory coupled to the gravitational sector through a phantom scalar field. We adopt a spherically symmetric background in (3+1) dimensions and consider two types of systems: electrically and magnetically charged. In the first case, we consider the generalized Ellis–Bronnikov model, in which we fix the power of the kinetic term in the k-essence Lagrangian function to $n=1/2$ and take the parameter $m\ge 2$, which acts as a generalization factor for the geometry of the wormhole area function. From this, we obtained the expression for the scalar field, the potential, and the associated electromagnetic functions for any values of the parameter $m\geq{2}$. In the second and third models, we consider the scenario of two wormholes that are structured according to the adjustment of the parameters that define the metric component associated with the area function $\Sigma^2$ (the $g_{22}$ component of the line element), and in both cases we adopt $n=1/2$. We show that the violation of the null energy conditions is conditioned by the parameters of the area function. Finally, we studied the linear stability of the models through the behavior of a test scalar field using both the WKB method and the time-domain evolution method.
\noindent 
\end{abstract}
	
	
\maketitle
	
\section{Introduction}\label{sec1}

Singular black hole solutions emerge naturally in the context of General Relativity (GR), being associated with the gravitational collapse that characterizes the final phase of the evolution of massive stars \cite{INTRO1,INTRO2,INTRO3}. In the 1970s, Bardeen proposed the first regular black hole solution \cite{INTRO4}, that is, a singularity free black hole. However, the spacetime he introduced did not satisfy the field equations in the vacuum regime. It was only around the 2000s that Beato and Garcia \cite{INTRO5} determined the content of matter compatible with this geometry, linking it to a class of nonlinear electrodynamic (NED) theories, in which the regularization parameter acquires the physical interpretation of the magnetic charge of a monopole. From the regular solution proposed by Bardeen, several other models of regular black holes were developed, contributing significantly to the advancement of theoretical understanding of these objects \cite{INTRO6,INTRO7,INTRO8,INTRO9,INTRO10,INTRO11,INTRO12,INTRO13,INTRO14,INTRO15,INTRO16,INTRO17,INTRO18}.

In the context of regular solutions, Simpson and Visser recently introduced a new class of geometries called black-bounce \cite{matt}. In general, the starting point is the static and spherically symmetric Schwarzschild spacetime, to which the regularization procedure $x^2\to {x^2+a^2}$ is applied, ensuring the regularity of the metric throughout all spatial domain. The parameter $a^2$ can be interpreted as characterizing the throat of a wormhole. Depending on how the parameter $a$ is related to the black hole mass $M$, we can have some types of possible configuration, as follows: with $a>2M$, we have a bidirectional traversable wormhole, $a=2M$, a unidirectional wormhole, and with the throat located at the origin $x=0$ and $a<2M$ we have a regular black hole with two symmetric horizons.
The regularized metric does not represent a vacuum configuration. Thus, in subsequent works, some authors obtained the content of matter responsible for maintaining the geometric structure of this spacetime, which was composed of the combination of non-linear electrodynamics and phantom scalar field \cite{INTRO19, INTRO20,Rodrigues:2023vtm}. As in the case of standard regular black holes, black-bounces can also be described by electrically charged sources \cite{Alencar:2024yvh}. In addition to spherically symmetric and static spacetimes, new black bounce models have been explored, for example, in the context of black strings \cite{INTRO21,INTRO22,INTRO23,INTRO24}, modified gravity \cite{INTRO25,INTRO25A,INTRO25B,INTRO25C}, stationary background \cite{INTRO26,INTRO27}, teleparallel gravity \cite{INTRO28} and conformal gravity \cite{INTRO29}. Black bounce solutions and their variations were also explored in the study of gravitational deflection for both light and massive particles \cite{INTRO30,INTRO31,INTRO32,INTRO33,INTRO34,INTRO35,INTRO36,INTRO37}, scattering \cite{INTRO38}, tidal forces \cite{Silva:2026mlo,Crispim:2025cql}, quasinormal modes and gravitational echoes \cite{INTRO39}.

The k-essence models, originally formulated to describe the dynamics of inflationary cosmology and later generalized to account for the accelerated expansion of the Universe, incorporate nonstandard kinetic terms for scalar fields coupled to gravity \cite{INTRO40,INTRO41}. In the context of compact objects, such as black holes and wormholes, k-essence has been used to propose exact solutions with intriguing properties, including configurations that challenge traditional notions of event horizons and singularities. Recently, in the work \cite{INTRO42}, the authors began the study by analyzing the k-essence field to describe cosmological evolution and then applied the regulation process to investigate wormhole models in $f(R)$. The model in question, which when analyzed in the cosmological scenario using only GR required the presence of the phantom scalar, when analyzed from the point of view of the $f(R)$ theory, this restriction ceases to exist. Still within the context of compact objects, the authors in \cite{INTRO43,INTRO44} analyze the spacetime of the Ellis–Bronnikov (EB) wormhole within the theories of k-essence and Rastall, as well as its stability conditions. From this perspective, in the present work we consider obtaining matter sources for three wormhole models for a k-essence field power that could provide analytical expressions for electrically and magnetically charged systems.

The paper is structured as follows: Section \ref{sec2} establishes the theoretical basis of the k-essence model and includes the derivation of the equations of motion for the electrically and magnetically charged system. In Sections \ref{sec3}, \ref{sec4} and \ref{sec5}, we apply the methodology to three specific models and derive the physical quantities for the k-essence configurations. In Section \ref{sec6}, we present a general derivation of the energy conditions and discuss their restrictions. Finally, the conclusions are presented in Section \ref{sec7}.

\section{General relations}\label{sec2}

The k-essence theories are characterized by the presence of scalar fields whose kinetic terms are introduced in a non-canonical way and have been recently explored in the context of wormhole solutions \cite{CDJM1, CDJM2,CDJM3}. Now, we will introduce the action representing gravity coupled minimally to the k-essence field and the NED. Therefore, consider the model described by the action below:
\begin{equation}\label{Lagran}
S=\int{d^4}{x}\sqrt{-g}[R-F(X,\phi) + L(f)]\,,
\end{equation}
where $R$ is the Ricci scalar, $X=\eta\phi_{;\rho}\phi^{;\rho}$ denotes the kinetic term, and $L(f)$ represents the contribution from electromagnetism, where $f=\frac{H_{\mu\nu}H^{\mu\nu}}{4}$, with $H_{\mu\nu}=\partial_\mu{A_\nu}-\partial_\nu{A_\mu}$ being the electromagnetic tensor, and $A_\mu$ being the four-dimensional vector potential. k-essence models can include a potential term and non-trivial couplings. Even in these cases, the scalar sector is typically minimally coupled to gravity. The parameter $\eta=\pm 1$ is introduced to avoid imaginary terms in the kinetic expression $X$. By selecting different forms for the function $F(X,\phi)$, k-essence theories can describe both phantom \cite{phan1, phan2, phan3, phan4} and standard scalar fields.

By varying the above action Eq. (\ref{Lagran}) with respect to the fields and the metric tensor, we obtain the following equations of motion:
\begin{eqnarray}\label{1}
 G_{\mu\nu}=T^{\phi}_{\mu\nu} + T^{EM}_{\mu\nu}, \\\label{2} 
\eta\nabla_\alpha\left(F_X\phi^{\alpha}\right)-\frac{1}{2}F_\phi=0, \\\label{3}
\nabla_\mu\left[L_{f}H^{\mu\nu}\right]=0,
\end{eqnarray} where $G_{\mu\nu}$ is the Einstein tensor, $T^{\phi}_{\mu\nu}$ and $T^{EM}_{\mu\nu}$ are the stress-energy tensors of the scalar field $\phi$ and the electromagnetic field, respectively, $F_X=\frac{\partial{F}}{\partial{X}}$, $F_\phi=\frac{\partial{F}}{\partial\phi}$, $\phi_\mu=\partial_\mu\phi$, and $L_f=\frac{\partial{L}}{\partial{f}}$.

The energy-momentum tensor for each of the fields is defined by:
\begin{eqnarray}\label{4}
T^{\phi}_{\mu\nu}= -\frac{F}{2}g_{\mu\nu} + \eta{F_{X}}\nabla_\mu{\phi}\nabla_\nu{\phi}, \\\label{5}
T^{EM}_{\mu\nu}= \frac{L(f)}{2}g_{\mu\nu} -\frac{L_f}{2}{H_\mu}^\alpha{H_{\nu\alpha}}.
\end{eqnarray}

The line element representing the most general spherically symmetric and static spacetime takes the form:
\begin{equation}\label{6}
ds^2=e^{2\gamma\left(u\right)}dt^2-e^{2\alpha\left(u\right)}du^2-e^{2\beta\left(u\right)}d\Omega^2,
\end{equation} where $u$ is an arbitrary radial coordinate and $d\Omega^2 = d\theta^2 + \sin^2\theta  d\varphi^2$ is the area element. Since our spacetime is spherically symmetric and static, we can assume that the scalar potential is a function only of the radial coordinate, $\phi = \phi\left(u\right)$.

\subsection{Magnetic case}\label{sec2A}

For our purposes, we are only interested in possible magnetically charged solutions. Thus, the non-zero component of the Maxwell-Faraday tensor is given by
\begin{equation}\label{7}
    H_{23}= q_{m}\sin\theta,
\end{equation} where the electromagnetic scalar is defined by
\begin{equation}\label{8}
    f(u)=\frac{1}{4}H^{\mu\nu}H_{\mu\nu}=\frac{q^2_m}{2e^{4\beta(u)}},
\end{equation} with $q_m$ being the magnetic charge.

Thus, we write the general equations of motion, which are the same as those contained in Refs. \cite{CDJM1, CDJM2,CDJM3, INTRO44}. However, they are now modified by NED. It is assumed that the function $X = -\eta e^{-2\alpha} (\phi')^2$ is positive, which implies that $\eta = -1$. The choice of a positive $\eta=-1$ is to ensure that we will not have physical quantities with imaginary contributions. As a result, the equations of motion take the form:

\begin{eqnarray}\label{9}
2\left(F_X{e^{-\alpha+2\beta+\gamma}}\phi'\right)' - {e^{\alpha+2\beta+\gamma}}F_\phi=0, \\\label{10}
{\gamma}'' + {\gamma}'\left(2{\beta}'+ {\gamma}'-{\alpha}'\right)-\frac{e^{2\alpha}}{2}\left(F-XF_X\right) + \frac{e^{2\alpha}}{2}\left[L(f)-\frac{q^2_{m}L_f}{e^{4\beta}}\right]=0, \\ \label{11}
-e^{2\alpha-2\beta} + {\beta}'' +{\beta}'\left(2{\beta}'+ {\gamma}'-{\alpha}'\right) -\frac{e^{2\alpha}}{2}\left[F-XF_X-L(f)\right]=0, \\ \label{12}
-e^{-2\beta} + e^{-2\alpha}{\beta}'\left({\beta}'+2{\gamma}'\right) -\frac{F}{2} + XF_X + \frac{L(f)}{2}=0.
\end{eqnarray}

The notation used here follows that used in reference \cite{INTRO44}. The following coordinate transformation is defined: $u = x$, and the \textit{quasi-global} gauge $\alpha(u) + \gamma(u) = 0$ is employed. As a result, the line element in Eq. (\ref{6}) can be expressed in the following form:

\begin{equation}
ds^2= A\left(x\right)dt^2- \frac{dx^2}{A\left(x\right)} - \Sigma^2\left(x\right)d\Omega^2,\label{line_x}
\end{equation}
where the metric functions are defined as $A(x) = e^{2\gamma} = e^{-2\alpha}$ and $e^\beta = \Sigma(x)$. The equations of motion defined in Eqs. (\ref{9}-\ref{12}) can be rewritten in the new coordinates. Combining Eqs. (\ref{10}-\ref{12}), we get the following expressions:
\begin{eqnarray}\label{14}
2A\frac{{\Sigma}''}{\Sigma} - XF_X =0, \\\label{15}
{A}''\Sigma^2 - A\left(\Sigma^2\right)''+ 2 -\frac{q^2_m{L_f}}{\Sigma^2} =0,
\end{eqnarray} where the primes now represent derivatives with respect to $x$.

The two remaining equations, Eq. (\ref{9}) and Eq. (\ref{12}), are rewritten in the new coordinates as
\begin{eqnarray}\label{16}
2\left(F_X{A\Sigma^2}\phi'\right)' - \Sigma^2F_\phi = 0, \\\label{17}
\frac{1}{\Sigma^2}\left(-1 + A'\Sigma'\Sigma + A{\Sigma'}^2\right) -\frac{F}{2} + XF_X + \frac{L(f)}{2} = 0.
\end{eqnarray}

It has been established in previous works \cite{CDJM1, CDJM2,CDJM3} that pursuing black-bounce solutions solely with the kinetic term of the k-essence function is not mathematically consistent. Therefore, when constructing these new solutions, we must incorporate a scalar potential given by 
\begin{equation}\label{F}
F(X) = F_0 X^n - 2V(\phi),
\end{equation}
where $F_0$ is a constant, $n$ is a real number and $V(\phi)$ is the potential function. Using Eq. (\ref{14}), we can derive a general expression for the scalar field that depends on both the angular metric function $\Sigma(x)$ and the metric function $A(x)$, unlike most studies, where the scalar field depends solely on the angular function and its derivatives \cite{INTRO44, PRL, Rodrigues:2023vtm}.

\subsection{Electric case}\label{sec2B}

In this section, we will deal with the case where the solution is electrically charged and therefore we must consider the antisymmetric property of the Maxwell-Faraday tensor $H_{01}=-H_{10}$. Thus, using the equation of motion Eq. (\ref{3}) together with the line element Eq. (\ref{line_x}), we find
\begin{equation}\label{18}
H^{10}=\frac{q_e}{L_f{\Sigma}^2},
\end{equation}
resulting in the following expression for the electromagnetic scalar
\begin{equation}\label{19}
f=\frac{1}{4}H^{\mu\nu}H_{\mu\nu}= -\frac{{q^2_e}}{2L^2_f{\Sigma^4}},
\end{equation} being $q_e$ the electric charge.

The two equations of motion that are altered due to the presence of electrically charged matter content are now written as:
\begin{eqnarray}\label{20}
\frac{1}{\Sigma^2}\left[-1+ A{\Sigma'}^2+ \Sigma{A'}{\Sigma'}+A{\Sigma}{\Sigma}''\right] -\frac{F}{2} + \frac{XF_X}{2} +\frac{L(f)}{2} +\frac{q^2_e}{2L_f{\Sigma^4}}&=&0, \\\label{21}
{A''}\Sigma + 2{A'}{\Sigma'} + \Sigma(XF_X-F) + \Sigma{L(f)}&=&0.
\end{eqnarray}

To obtain the first equation (\ref{20}) we have to consider the difference between the components $(11)$ and $(00)$ in Einstein's equations (\ref{1}) and perform some algebraic manipulations. Likewise, to find the second equation of motion (\ref{21}) we just need to add the components from $(00)$ to $(22)$ in Einstein's equations (\ref{1}) and manipulate it.

The electromagnetic expressions $L(f)$ and $L_f$ are obtained independently through Einstein's equations for both the electric and magnetic cases and can be related through a consistency relation below:
\begin{equation}\label{22}
L_f-\frac{\partial{L}}{\partial{f}}=L_f- \left(\frac{\partial{L}}{\partial{x}}\right)\left(\frac{\partial{f}}{\partial{x}}\right)^{-1}=0.
\end{equation}

Electrically charged solutions do not even allow for an analytical expression for electromagnetic functions, and therefore it is often useful to consider an auxiliary field $P_{\mu\nu}=L_f{H_{\mu\nu}}$. Associated with this field, we have the invariant $P$, given by:
\begin{equation}\label{23}
P=\frac{1}{4}P^{\mu\nu}P_{\mu\nu}= -\frac{q^2_e}{2\Sigma^4}.
\end{equation} The above invariant is generally easier to invert than the scalar Eq. (\ref{19}). The scalar $P$ turns out to be much simpler to work with, since $L(P)$ is usually a function that can be written analytically in most cases. Typically, the scalar $f(x)$ is not analytically invertible because the Lagrangian $L(f)$ is multivalued.

\section{FIRST MODEL}\label{sec3}

In this section, we will study the generalized Ellis-Bronnikov (GEB) spacetime with the aim of obtaining the sources of matter for scenarios where the system is magnetically charged and electrically charged \cite{phan1,phan2,EB3}. This generalization arises by introducing a dimensionless parameter $m$ in the angular sector of the EB geometry. By varying $m$, one interpolates between the standard EB wormhole (for $m=2$) and configurations whose angular part becomes increasingly cylindrical as $m$ increases, thus generating a continuous family of wormhole throats with different curvatures and shapes.

 This model was investigated in obtaining its respective matter sources for the canonical case in \cite{EB3}, in the study of particle dynamics \cite{EB5}, wormhole solution in the canonical Lagrangian described in a modified gravity theory $f(R)$ \cite{EB6}, wormhole described in graphene \cite{EB7}, in the study of gravitational deflection \cite{EB8} and quantum systems \cite{EB9,EB10,EB11}.

\begin{figure}[htb!]
\centering  
	\mbox{
	\subfigure[]{\label{Sigma1a}
	{\includegraphics[width=0.45\linewidth]{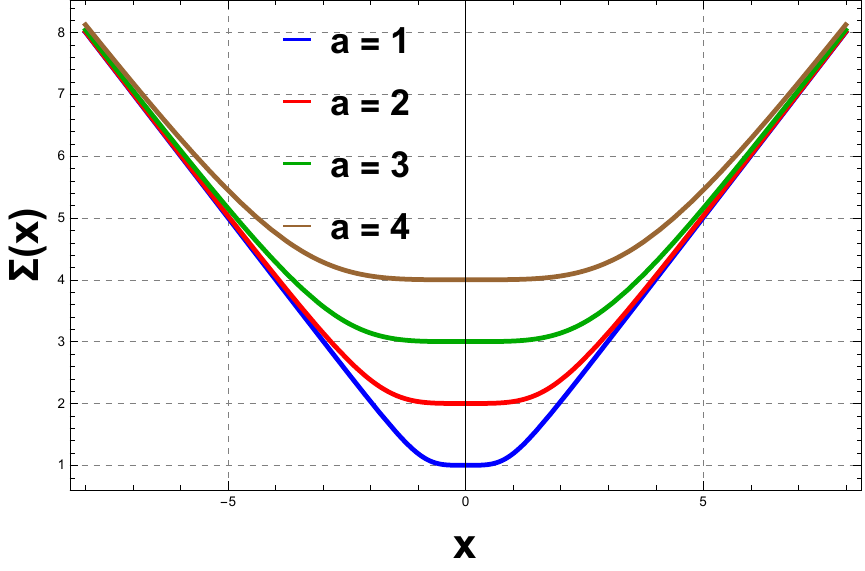}}}
	\subfigure[]{\label{Sigma1m}
	{\includegraphics[width=0.45\linewidth]{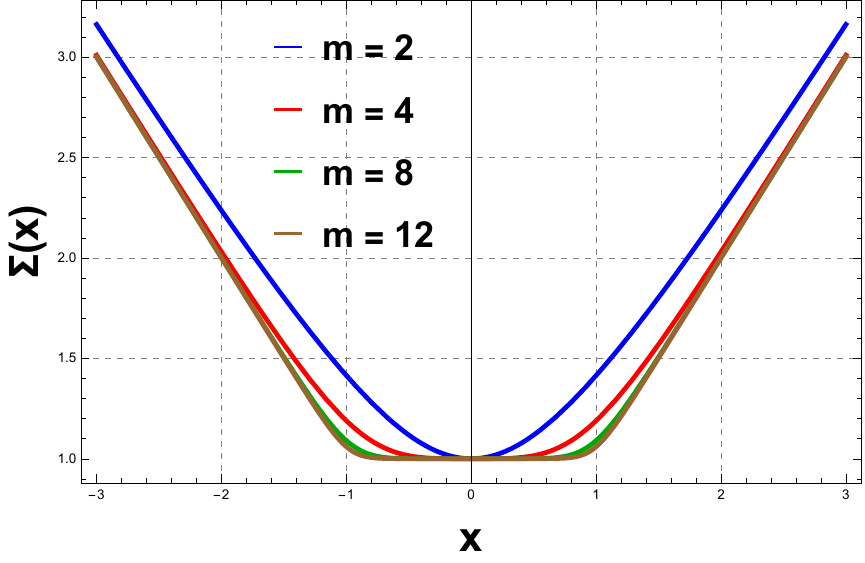}}}}
\caption{Behavior of $\Sigma$ of the GEB spacetime \eqref{line1} as a function of the coordinate $x$. In panel~(a), we fix $m=4$ and vary the values of $a$, while in panel~(b) we fix $a=1$ and vary the value of $m$.}
\label{FIGSigmaMOD1}
\end{figure}

Such spacetime is described by the line element Eq. (\ref{line_x}) with $A(x)=1$ and has a generalized area function defined as $\Sigma(x)=(x^m+a^m)^{1/m}$ where $a$ represents the radius of the wormhole throat and the free parameter $m\geq{2}$, for the particular case $m=2$ the EB wormhole is recovered. The line element is written as
\begin{equation}\label{line1}
    ds^2= dt^2-dx^2 - \left(x^m+a^m\right)^{2/m}d\Omega^2.
\end{equation}
In Fig.~\ref{FIGSigmaMOD1}, we show the behavior of the function $\Sigma$ for different values of the parameters. We observe that changing the parameter $a$, panel~\ref{Sigma1a}, increases the size of the throat radius, which is located at $x=0$. This is a direct consequence of the wormhole throat being located at the minimum point of $\Sigma$. From panel~\ref{Sigma1m}, we see that varying $m$ does not modify the location of the throat but instead changes the shape of the wormhole geometry. Since there is always one minimum point, the parameters $a$ and $m$ do not modify the number of throats that this wormhole can possess.

In the quantities defined in the following, we consider $\eta=-1$.

\subsection{Magnetic case}\label{sec31}
For this case already explained in section (\ref{sec2A}) we are considering the magnetically charged case and therefore for the generalized area function mentioned above and considering in equation (\ref{14}) the power-type form for the k-essence field, we obtain the general expression for the scalar field:
\begin{eqnarray}\label{eq311}
\phi(x)&=& 2^{\frac{2n+1}{2n}}\left(\frac{1}{F_0{n}}\right)^{\frac{1}{2n}}\left(\frac{nx}{2(n-1)+m}\right)\left[1+\left(\frac{x}{a}\right)^m\right]\left[\frac{(m-1)a^m{x^{m-2}}}{(a^m+x^m)^2}\right]^{1/2n} \\\nonumber
&\times &{_{2}}F_{1}\left[1,-\left(\frac{m+2(1-n(m+1))}{2mn}\right),\left(1+\frac{m+2(n-1)}{2mn}\right),-\left(\frac{x}{a}\right)^m\right],
\end{eqnarray}
where the above parameters are defined as $m$ belongs to the generalized area function, $n$ represents the power of the k-essence field, $F_0$ a free parameter that multiplies the kinetic term $X$ and ${_{2}}F_{1}$ the hypergeometric function. The above scalar field is recovered for the standard case described in the literature $F_0=n=1$ and $m=2$, thus, $\phi(x)=\sqrt{2}\arctan\left(\frac{x}{a}\right)$.

Using Eq. (\ref{16}) making use of the generalized area function and the scalar field, we find the expression referring to the scalar potential:
\begin{eqnarray}\label{eq312}
V(x)&=& \frac{(m-1)x^{m-2}\left[a^m\left(4n+m^2(1-2n)\right)-2(m-2)nx^m\right]}{m^2n(x^m+a^m)^2}    \\\nonumber
&-&\frac{4(m-1)x^{m-2}}{m^2a^m} {_{2}}F_{1}\left[1,1-\frac{2}{m},2-\frac{2}{m},-\left(\frac{x}{a}\right)^m\right].
\end{eqnarray}

In a similar way to what was developed in the quantities above, we can use Eqs. (\ref{15}) and (\ref{17}) to obtain the electromagnetic functions:
\begin{equation}\label{eq313}
L_f(x)= \frac{(a^m+x^m)^{2/m}}{q^2_m}\left[2-2(x^m+a^m(m-1))x^{m-2}(a^m+x^m)^{\frac{2}{m}-2}\right],
\end{equation}
\begin{equation}\label{eq314}
L(x)=  2\left[\frac{(m-2)x^{2m-2}}{m(a^m+x^m)^2} + \frac{4(1-m)x^{m-2}}{m^2(a^m+x^m)}+(a^m+x^m)^{-2/m}\right] 
+\frac{8(m-1)x^{m-2}}{m^2a^m}{_{2}}F_{1}\left[1,1-\frac{2}{m},2-\frac{2}{m},-\left(\frac{x}{a}\right)^m\right].
\end{equation}

We can use Eq. (\ref{8}) to invert the Lagrangian function above and write it in terms of the invariant $f=\frac{q^2_m}{2(a^m+x^m)^{4/m}}$. Thus, we have:
\begin{eqnarray}\label{eq315}
L(f)&=&  \frac{2(m-2)}{ma^2}\left(\frac{s}{f^{1/4}}\right)^{-2m}\left[\left(\frac{s}{f^{1/4}}\right)^m-1\right]^{\frac{2m-2}{m}} + \frac{8(1-m)}{m^2a^2}\left(\frac{s}{f^{1/4}}\right)^{-m}\left[\left(\frac{s}{f^{1/4}}\right)^m-1\right]^{\frac{m-2}{m}} \\\nonumber
&+& \frac{8(m-1)}{m^2a^2}\left[\left(\frac{s}{f^{1/4}}\right)^m-1\right]^{\frac{m-2}{m}}{_{2}}F_{1}\left[1,1-\frac{2}{m},2-\frac{2}{m},1-\left(\frac{s}{f^{1/4}}\right)^m\right] + \frac{2}{a^2}\left(\frac{s}{f^{1/4}}\right)^{-2},
\end{eqnarray} in which $s=\left(\frac{q^2_m}{2a^4}\right)^{1/4}$.

For simplicity, we can define an auxiliary variable $\xi=s/f^{1/4}=\Sigma/a$ and rewrite the Lagrangian above as
\begin{eqnarray}\label{eq316}
L(\xi)&=&\frac{8(m-1)\left(\xi^m-1\right)^{\frac{m-2}{m}}}{m^2a^2} {_{2}}F_{1}\left[1,1-\frac{2}{m},2-\frac{2}{m},1-\xi^m\right] + \frac{2}{a^2\xi^2} \\\nonumber
&+& \frac{2(m-2)\left(\xi^m-1\right)^{\frac{2m-2}{m}}}{ma^2\xi^{2m}} + \frac{8(1-m)\left(\xi^m-1\right)^{\frac{m-2}{m}}}{m^2a^2\xi^{m}}.
\end{eqnarray}
As already discussed in Refs. \cite{CDJM3,CDJM4}, the electromagnetic functions do not depend on the k-essence theory parameter as a non-canonical kinetic term $n$. This condition can be demonstrated by Eqs. (\ref{eq313}) and (\ref{eq314}) where the electromagnetic functions depend only on the parameter related to the generalized area function and the radius of the wormhole throat. 
  
\begin{figure}[htb!]
\centering  
	\mbox{
	\subfigure[]{\label{EBCAMPO1}
	{\includegraphics[width=0.3\linewidth]{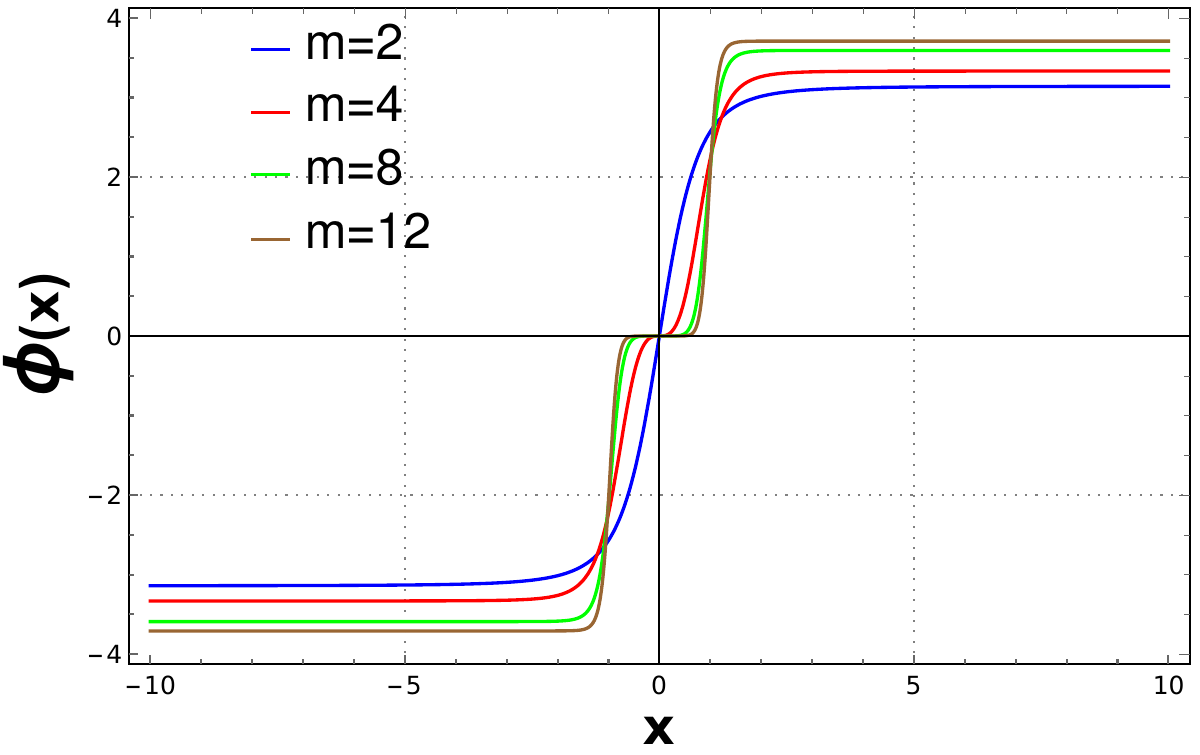}}}\qquad
	\subfigure[]{\label{EBPOT1}
	{\includegraphics[width=0.3\linewidth]{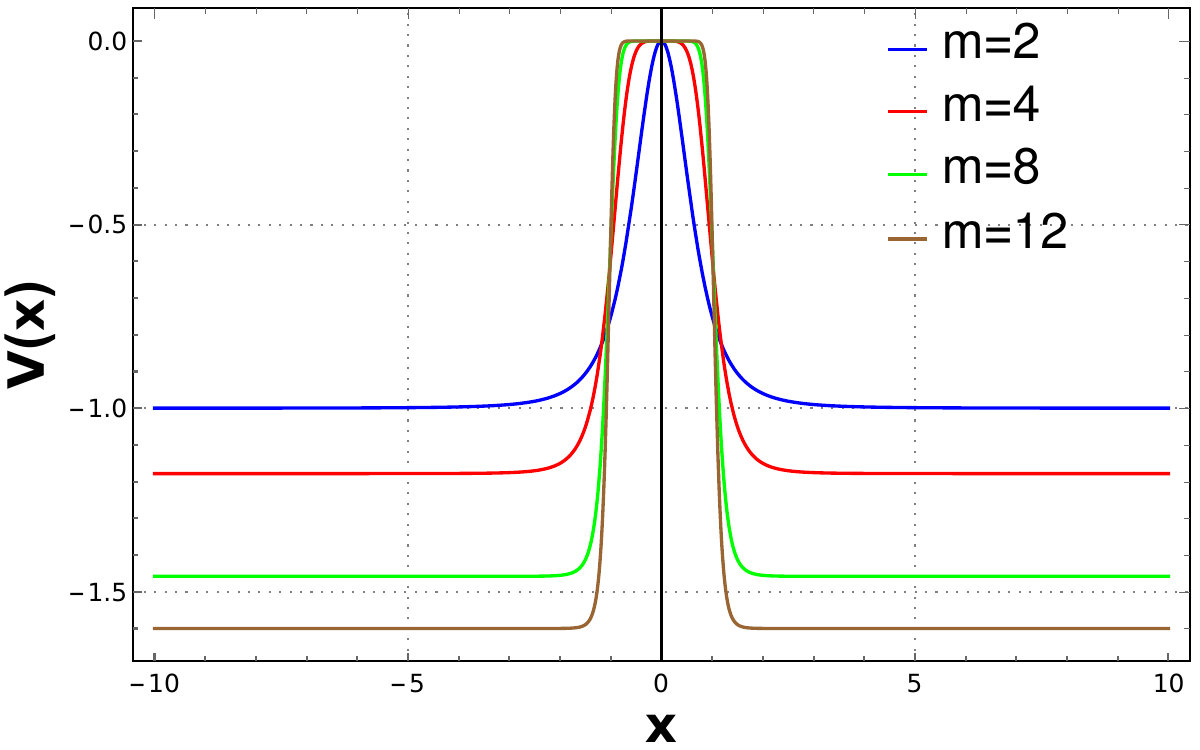}}}\qquad
    \subfigure[]{\label{EBLAG1}
	{\includegraphics[width=0.3\linewidth]{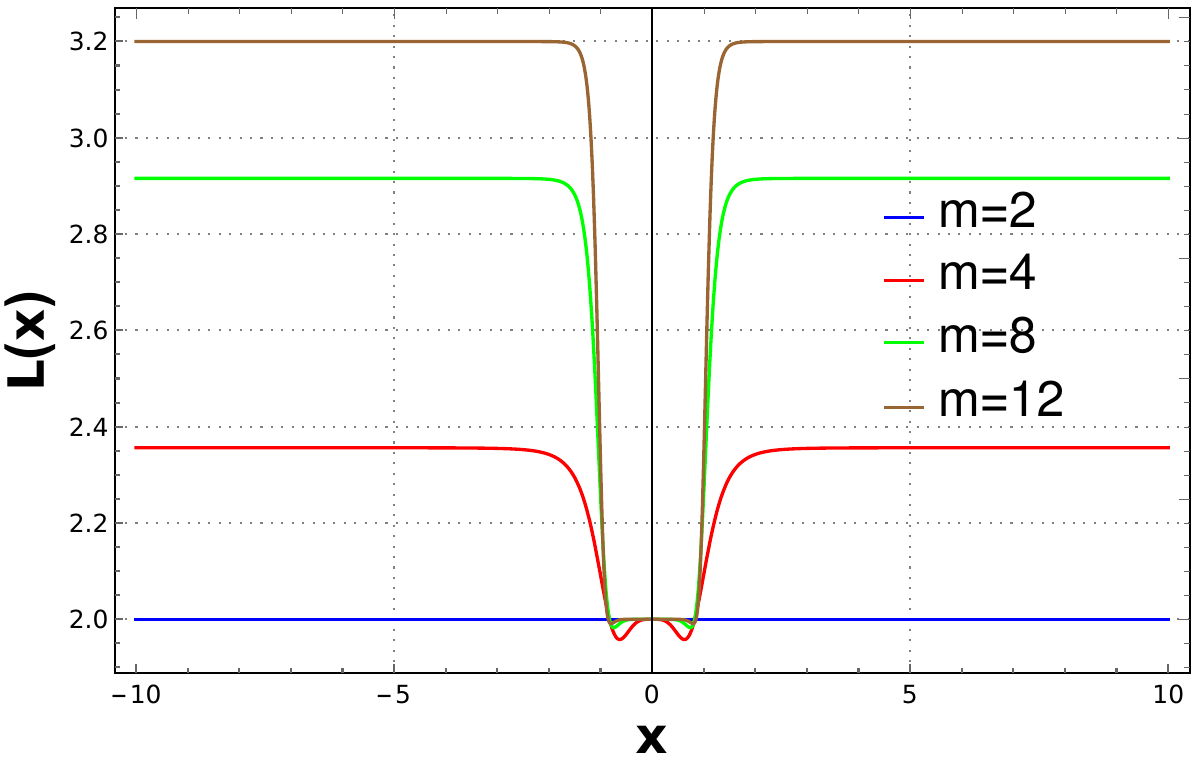}}}}
\caption{In the figures above the following values were set $n=1/2$ and $a=F_0=1$. In a) we have the scalar field, b) the potential and in c) the Lagrangian.}
\label{FIG1MOD1}
\end{figure}

In Fig. \ref{FIG1MOD1}, we have the graphical representation for the scalar field Eq. (\ref{eq311}), the potential Eq. (\ref{eq312}), and the Lagrangian function Eq. (\ref{eq314}) varying  as a function of $x$ for some fixed values for the power of the generalized area function $m$. For this combination of matter between NED and phantom scalar field to be sustained, it is necessary that $m>2$ this can be illustrated by the blue curve of Eq. (\ref{eq314}) where the Lagrangian function assumes a constant value and must be contrary to the same case of the potential Eq. (\ref{eq312}).

\subsection{Electric case}\label{sec32}

Still for the GEB model, we will construct the electromagnetic quantities related to an electrically charged solution. To do so, we will initially consider Eqs. (\ref{20}) and (\ref{21}) and then make use of the generalized area function, as well as the scalar field Eq. (\ref{eq311}) and potential Eq. (\ref{eq312}). Below are the electromagnetic functions for the scenario where the source of matter is electrically charged
\begin{eqnarray}\label{eq321}
L(x)&=& \frac{2(m-1)(m-2)x^{m-2}\left(2x^m+a^m(m+2)\right)}{m^2(a^m+x^m)^2} 
+ \frac{8(m-1)x^{m-2}}{m^2a^m}{_{2}}F_{1}\left[1,1-\frac{2}{m},2-\frac{2}{m},-\left(\frac{x}{a}\right)^m\right],\\
L_f(x)&=& \frac{q^2_e(a^m+x^m)^{-2/m}}{2\left[1-x^{m-2}(a^m+x^m)^{\frac{2-2m}{m}}\left(x^m+a^m(m-1)\right)\right]}.
\end{eqnarray}
As in the previous section, we can construct a Lagrangian functional as compact as possible by making use of the auxiliary variable Eq. (\ref{23}). Thus, we have that the Lagrangian Eq. (\ref{eq321}) becomes
\begin{eqnarray}\label{eq322}
L(P)&=& \frac{2(m-1)(m-2)}{m^2a^2}\left(-\frac{s_1}{P^{1/4}}\right)^{-2m}\left[-1+\left(-\frac{s_1}{P^{1/4}}\right)^m\right]^{\frac{m-2}{m}}\left[m+2\left(-\frac{s_1}{P^{1/4}}\right)^m\right] \\\nonumber 
&+& \frac{8(m-1)}{m^2a^2}\left[-1+\left(-\frac{s_1}{P^{1/4}}\right)^m\right]^{\frac{m-2}{m}}{_{2}}F_{1}\left[1,1-\frac{2}{m},2-\frac{2}{m},1-\left(-\frac{s_1}{P^{1/4}}\right)^m\right],
\end{eqnarray} in which $s_1=(\frac{q^2_e}{2a^4})^{1/4}$.

\begin{figure}[htb!]
\centering  
	\subfigure[]{\label{EBFX}
	{\includegraphics[width=.47\linewidth]{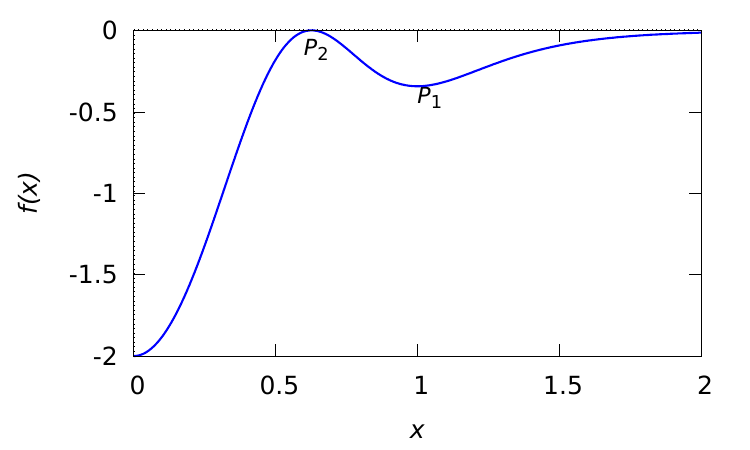}}}
	\subfigure[]{\label{EBFP}
	{\includegraphics[width=.47\linewidth]{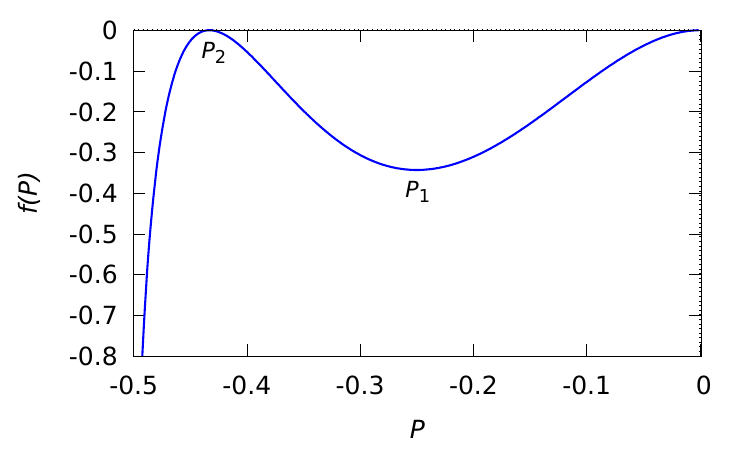}}}
    \subfigure[]{\label{EBLF}
	{\includegraphics[width=.47\linewidth]{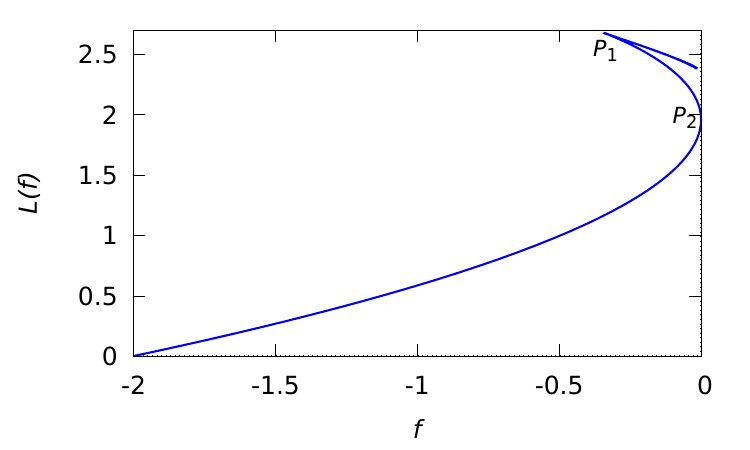}}}
\caption{In the figures above the following values were set $q_e=a=1$ and $m=4$. In a) we have the scalar $f(x)$, in b) the electromagnetic scalar as a function of the scalar $P$, and in c) the Lagrangian.}
\label{FIG2MOD1}
\end{figure}

In order to better understand the behavior of the electromagnetic Lagrangian for the electrically charged case, let us analyze Fig. \ref{FIG2MOD1}. From the behavior of the functions $f(x)$ or $f(P)$, it becomes clear that these functions cannot be inverted analytically, since they are multivalued. As a consequence, the electromagnetic Lagrangian $L(f)$ exhibits three different behaviors depending on the value of the coordinate $x$. The change in the behavior of the Lagrangian occurs at the points $P_1$ and $P_2$. While $P_2$ corresponds to a smoother transition, as $f=0$ at this point, $P_1$ represents a sharper change characterized by a cusp in the curve.

The expression referring to the electric field can be obtained from Eq.(\ref{18}) $H^{01}=-H^{10}=E(x)$, so we have that

\begin{equation}\label{eq323}
E(x)=  -\frac{2\left[1-x^{m-2}(a^m+x^m)^{\frac{2-2m}{m}}\left(x^m+a^m(m-1)\right)\right]}{q_e}.
\end{equation}
If we expand the electric field in regions where $x\to \infty$, we find that
\begin{equation}
    E(x\to\infty) \propto \frac{a^m}{x^m}.
\end{equation}
However, this result is not valid for $m=2$ since the electric field is always zero in this case. So, the electric field does not behave as the Coulomb field.

\section{SECOND MODEL}\label{sec4}

\begin{figure}[htb!]
\centering  
	\mbox{
	\subfigure[]{\label{Sigma2c3}
	{\includegraphics[width=0.32\linewidth]{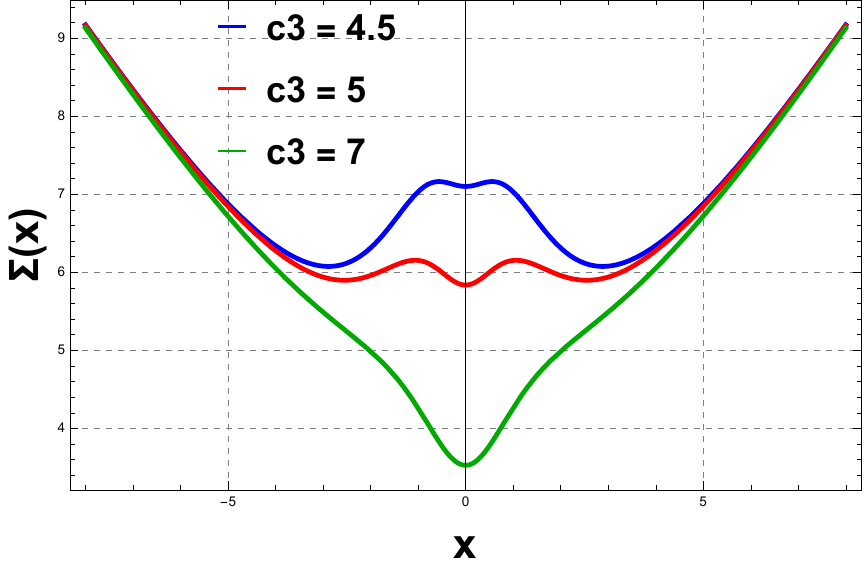}}}
	\subfigure[]{\label{Sigma2b}
	{\includegraphics[width=0.32\linewidth]{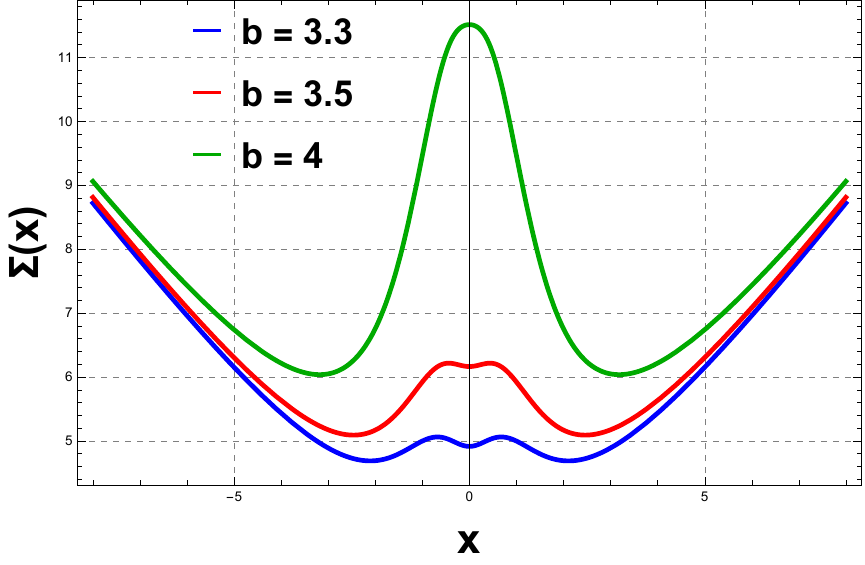}}}
    \subfigure[]{\label{Sigma2d}
	{\includegraphics[width=0.32\linewidth]{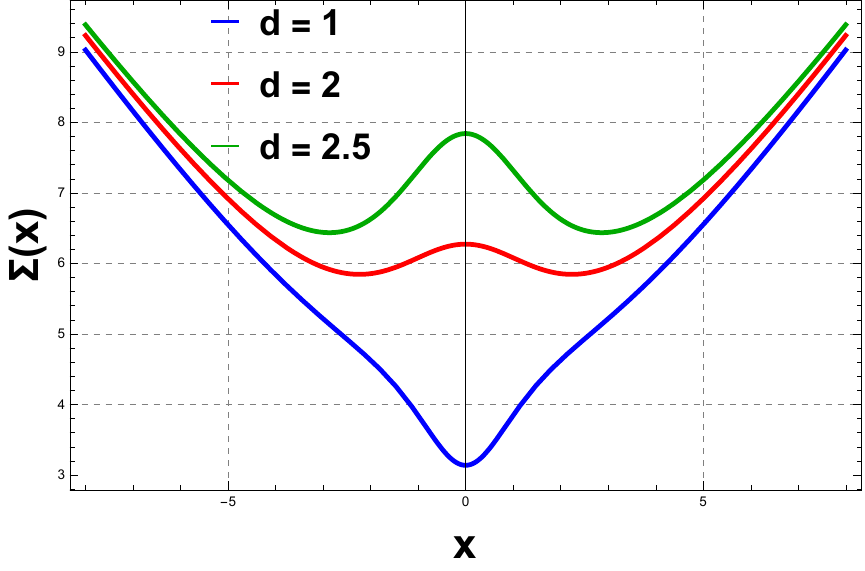}}}}
\caption{Behavior of $\Sigma$ of the spacetime \eqref{line2} as a function of the coordinate $x$. In panel~(a), we fix $b=4.2$, $d=1$ and vary the values of $c_3$. In panel~(b) we fix $c_3=3$, $d=0.8$ and vary the value of $b$. In panel~(c) we fix $c_3=7$, $b=4$ and vary the value of $d$.}
\label{FIGSigmaMOD2}
\end{figure}

In this section, we will study a wormhole model that was originally motivated by the new black-bounce solutions obtained by the local variation of the Hernandez–Misner–Sharp mass \cite{LOBO1} and then extended to the variation of the area function \cite{MR1CQG}. Thus, the area function is defined as $\Sigma^2(x)= \left(d^2+x^2\right)e^{\frac{b^2}{{c_3}+x^2}}$, where $d,b$, and ${c_3}$ are parameters that modify the structural shape of the wormhole. The line element is written as
\begin{equation}\label{line2}
    ds^2= dt^2- dx^2 - \left(d^2+x^2\right)e^{\frac{b^2}{{c_3}+x^2}}d\Omega^2.
\end{equation}
As discussed in \cite{MR1CQG}, the first and second derivatives with respect to the radial coordinate, for the case in which the parameters $b\neq{0}$ and $c_3=0$ have an exponential divergence, and therefore the curvature scalar will also be divergent, generating systems that have singularities. Thus, the relationship between these parameters is essential for us to generate new regular geometries, describing wormholes that encompass the presence of multiple throats and anti-throats. In Fig.~\ref{FIGSigmaMOD2}, we show the behavior of $\Sigma$ for the model~\eqref{line2}. It is noteworthy that changing the parameters modifies the number, location, and values of the maxima and minima points. In this way, the presence of throats and anti-throats is directly affected by the choice of parameters. Depending on the choice of parameters, this wormhole solution may also exhibit multiple stable and unstable photon orbits, which play an important role in the optical appearance of these solutions and in their stability.

Thus, in order to have an analytical solution for the physical quantities of interest, we will consider the configuration in which the k-essence field has power $n=1/2$ and the parameter $\eta=-1$, this for both the magnetically charged system and the electrically charged system.

\subsection{Magnetic case}\label{MOD2MG1}
For this case already explained in section (\ref{sec2A}) we are considering the magnetically charged system and, therefore, for the area function of interest mentioned above and considering in equation (\ref{14}) the power $n=1/2$ for the k-essence field, we obtain the general expression for the scalar field:

\begin{eqnarray}\label{CESCALARMOD2MAGN}
    \phi\left(x\right)&=& \frac{b^2 x \left(b^2 \left(\frac{8 c_3 x^2+3 x^4}{c_3^2}-3\right)-\frac{48 \left(c_3+x^2\right) \left(2 c_3-d^2+x^2\right)}{c_3-d^2}\right)}{12 \left(c_3+x^2\right){}^3}+\frac{b^2 \left(b^2 \left(c_3-d^2\right){}^2-16 c_3^2 \left(c_3+d^2\right)\right) \tan ^{-1}\left(\frac{x}{\sqrt{c_3}}\right)}{4 c_3^{5/2} \left(c_3-d^2\right){}^2} \\\nonumber
    &+&   \left(\frac{8 b^2 d}{\left(c_3-d^2\right){}^2}+\frac{2}{d}\right) \tan ^{-1}\left(\frac{x}{d}\right)+\frac{2 x}{d^2+x^2},
\end{eqnarray} in which we fix the parameters in the expression above as $F_0=1$ and $\eta=-1$. It is worth noting that in the limit where $b\to{0}$ the same scalar field is recovered in the k-essence theory $n=1/2$ and corresponds to the GEB model \cite{phan1,phan2} for $m=2$ analyzed in Section (\ref{sec31}).

Using Eq. (\ref{16}) making use of the generalized area function and the scalar field Eq. (\ref{CESCALARMOD2MAGN}), we define the expression referring to the scalar potential:
\begin{eqnarray}\label{POTMOD2}
   V(x)&=& \frac{6 b^2 \left(d^2 (b-d) (b+d)+c_3^2\right) \left(\log \left(d^2+x^2\right)-\log \left(c_3+x^2\right)\right)}{\left(d^2-c_3\right){}^4} + \frac{3 b^4 d^2+4 b^2 c_3 \left(c_3-d^2\right)}{\left(d^2-c_3\right){}^2 \left(c_3+x^2\right){}^2} \\\nonumber 
   &-&\frac{\left(c_3-d^2\right) \left(\frac{4 b^6 c_3 \left(d^2-c_3\right){}^3}{\left(c_3+x^2\right){}^5}-\frac{20 b^4 \left(d^3-c_3 d\right){}^2}{\left(c_3+x^2\right){}^3}+\frac{20 b^2 d^2 \left(d^2-c_3\right)}{d^2+x^2}+\frac{10 d^2 \left(d^2-c_3\right){}^3}{\left(d^2+x^2\right)^2}\right)}{10 \left(d^2-c_3\right){}^4} \\\nonumber 
   &-&\frac{b^2 \left(c_3-d^2\right) \left(b^2 \left(b^2-4 c_3\right) \left(c_3-d^2\right){}^3+4 \left(c_3+x^2\right){}^3 \left(d^2 \left(c_3-3 b^2\right)-3 c_3^2+2 d^4\right)\right)}{2 \left(d^2-c_3\right){}^4 \left(c_3+x^2\right){}^4}.
\end{eqnarray}

Similarly to what was developed above to determine the expressions referring to the scalar field and the potential, we can use Eqs. (\ref{15}) and (\ref{17}) to obtain the following electromagnetic functions:
\begin{eqnarray}\label{PRIMEMAGFUNCTIONMOD2}
    L_{f}(x)=\frac{\left(d^2+x^2\right) e^{\frac{b^2}{c_3+x^2}} \left(-\frac{2 b^2 \left(d^2+x^2\right) e^{\frac{b^2}{c_3+x^2}} \left(2 x^2 \left(b^2+c_3\right)-c_3^2+3 x^4\right)}{\left(c_3+x^2\right){}^4}+\frac{8 b^2 x^2 e^{\frac{b^2}{c_3+x^2}}}{\left(c_3+x^2\right){}^2}-2 e^{\frac{b^2}{c_3+x^2}}+2\right)}{q_m^2}, 
\end{eqnarray}

\begin{eqnarray} \label{SECUNDMAGFUNCTIONMOD2}
    L(x)&=& \frac{b^6 \left(c_3+5 x^2\right)}{5 \left(c_3+x^2\right){}^5}+\frac{2 \left(e^{-\frac{b^2}{c_3+x^2}}-1\right)}{d^2+x^2}-\frac{4 b^2 \left(3 x^4 \left(c_3+d^2\right)+6 c_3^2 d^2+x^2 \left(7 c_3 d^2+4 c_3^2+d^4\right)\right)}{\left(d^2-c_3\right){}^2 \left(c_3+x^2\right){}^2 \left(d^2+x^2\right)} \\\nonumber
  &+&  \frac{2 b^4 \left(c_3 \left(c_3 \left(c_3 \left(2 c_3+5 d^2+x^2\right)-d^4+23 d^2 x^2\right)-7 d^2 x^2 \left(d^2-3 x^2\right)\right)+d^2 x^2 \left(d^4-3 d^2 x^2+6 x^4\right)\right)}{\left(d^2-c_3\right){}^3 \left(c_3+x^2\right){}^4}  \\\nonumber
  &-&\frac{12 b^2 \left(d^2 (b-d) (b+d)+c_3^2\right) \left(\log \left(d^2+x^2\right)-\log \left(c_3+x^2\right)\right)}{\left(d^2-c_3\right){}^4}.
\end{eqnarray}

In Eqs. (\ref{PRIMEMAGFUNCTIONMOD2}) and (\ref{SECUNDMAGFUNCTIONMOD2}) we have the expressions referring to the electromagnetic quantities for the area function of interest and, as mentioned previously, these quantities do not depend on the power $n$ of the k-essence field. Furthermore, we can see, for both the scalar field, potential and electromagnetic quantities, that there is a validity restriction between the parameters of the area function $d^2-c_3>0$ and $c_3\neq{0}$, otherwise these quantities tend to diverge.

\begin{figure}[htb!]
\centering  
	\mbox{
	\subfigure[]{\label{CAMPOMOD21}
	{\includegraphics[width=0.3\linewidth]{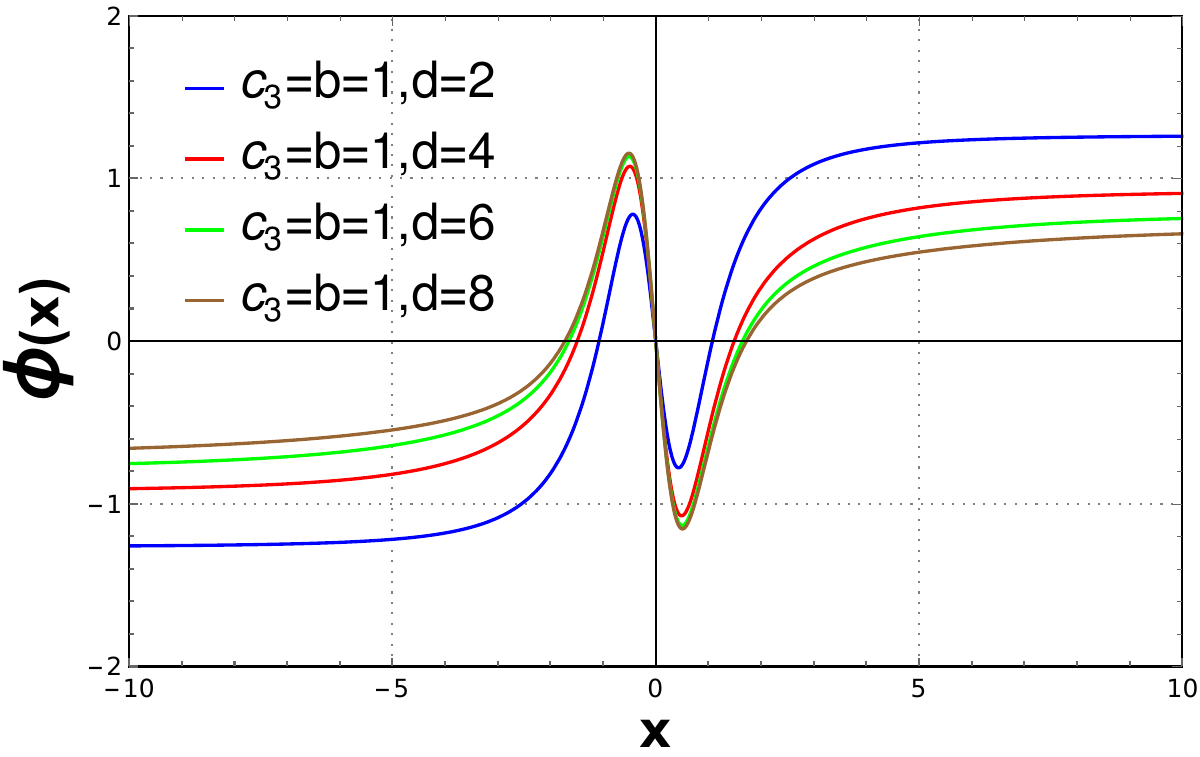}}}\qquad
	\subfigure[]{\label{CAMPOMOD22}
	{\includegraphics[width=0.3\linewidth]{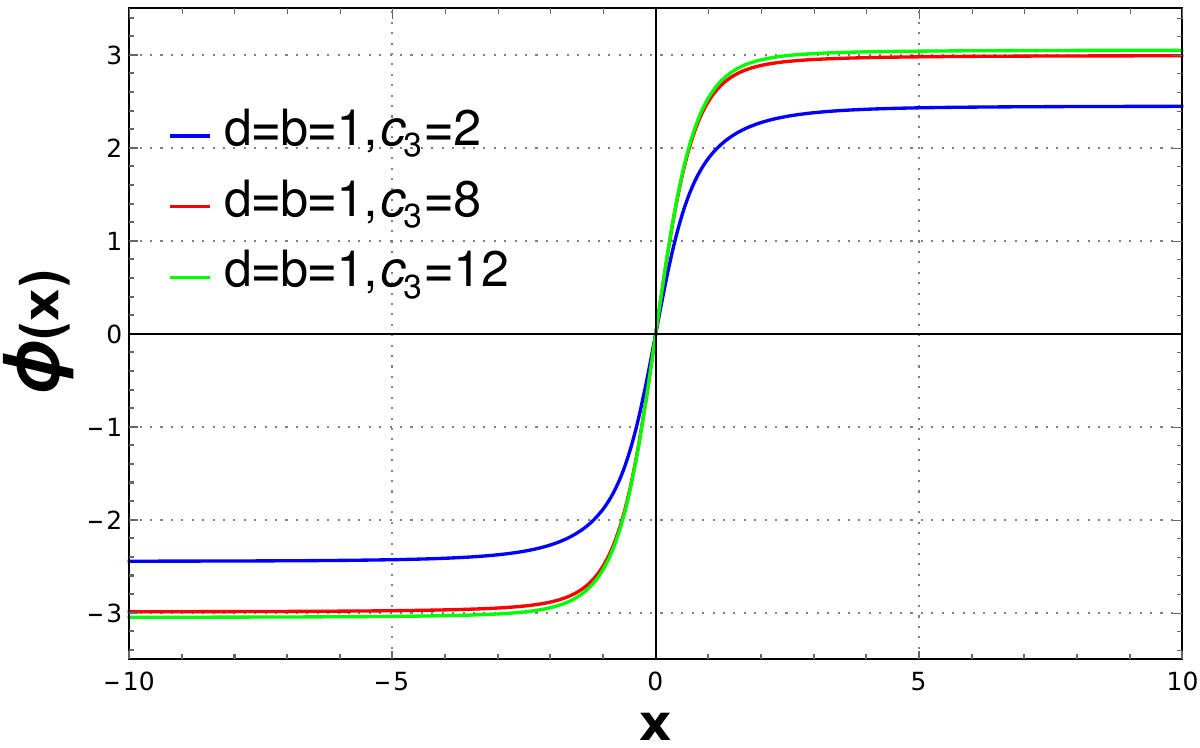}}}\qquad
    \subfigure[]{\label{CAMPOMOD23}
	{\includegraphics[width=0.3\linewidth]{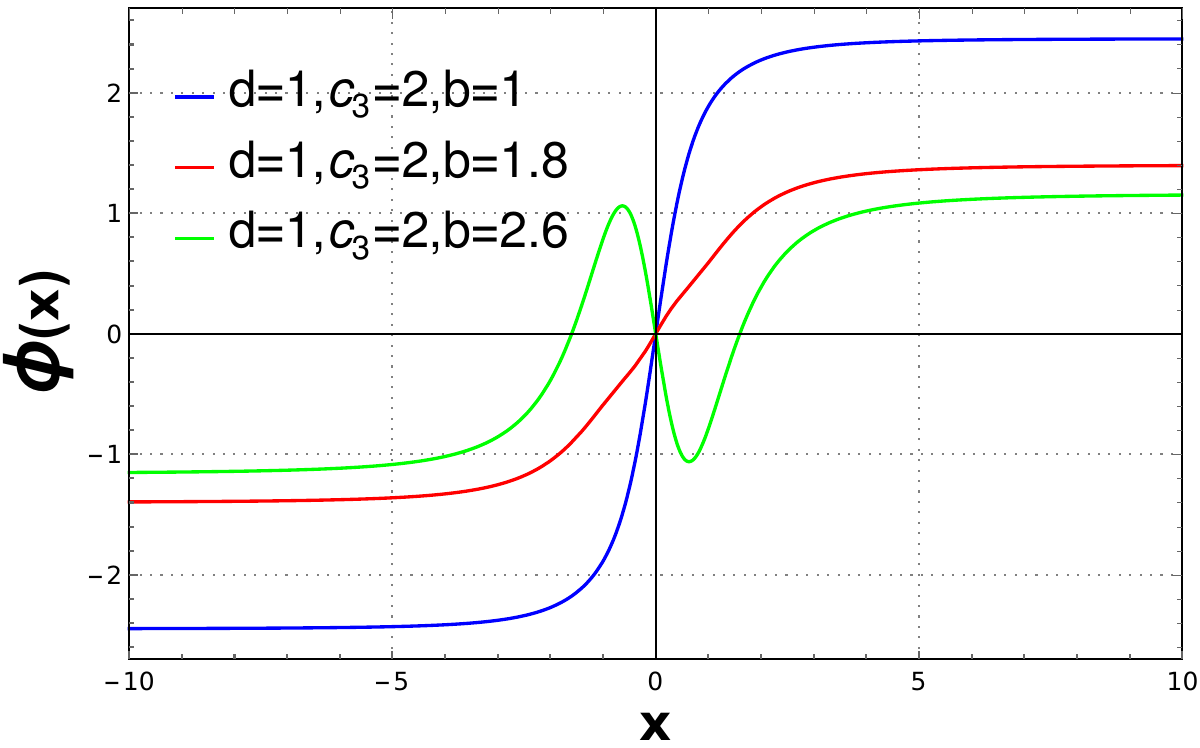}}}} \\
    \mbox{
	\subfigure[]{\label{POTEMOD21}
	{\includegraphics[width=0.3\linewidth]{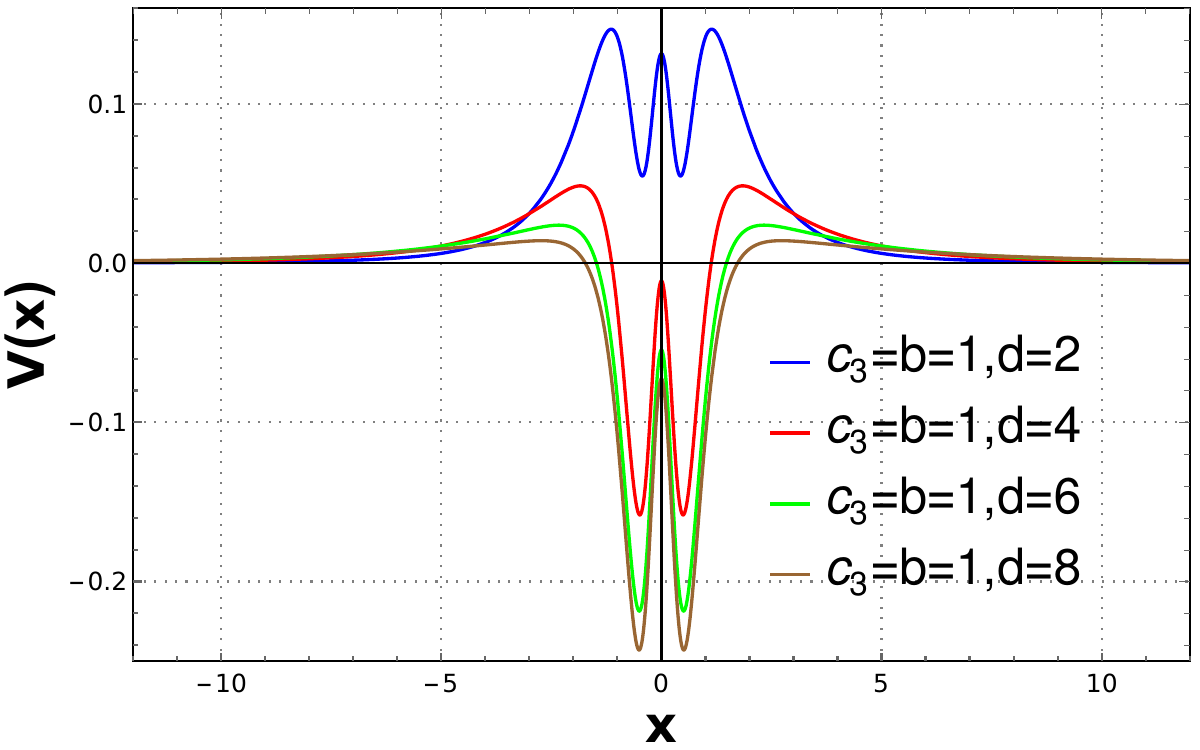}}}\qquad
	\subfigure[]{\label{POTEMOD22}
	{\includegraphics[width=0.3\linewidth]{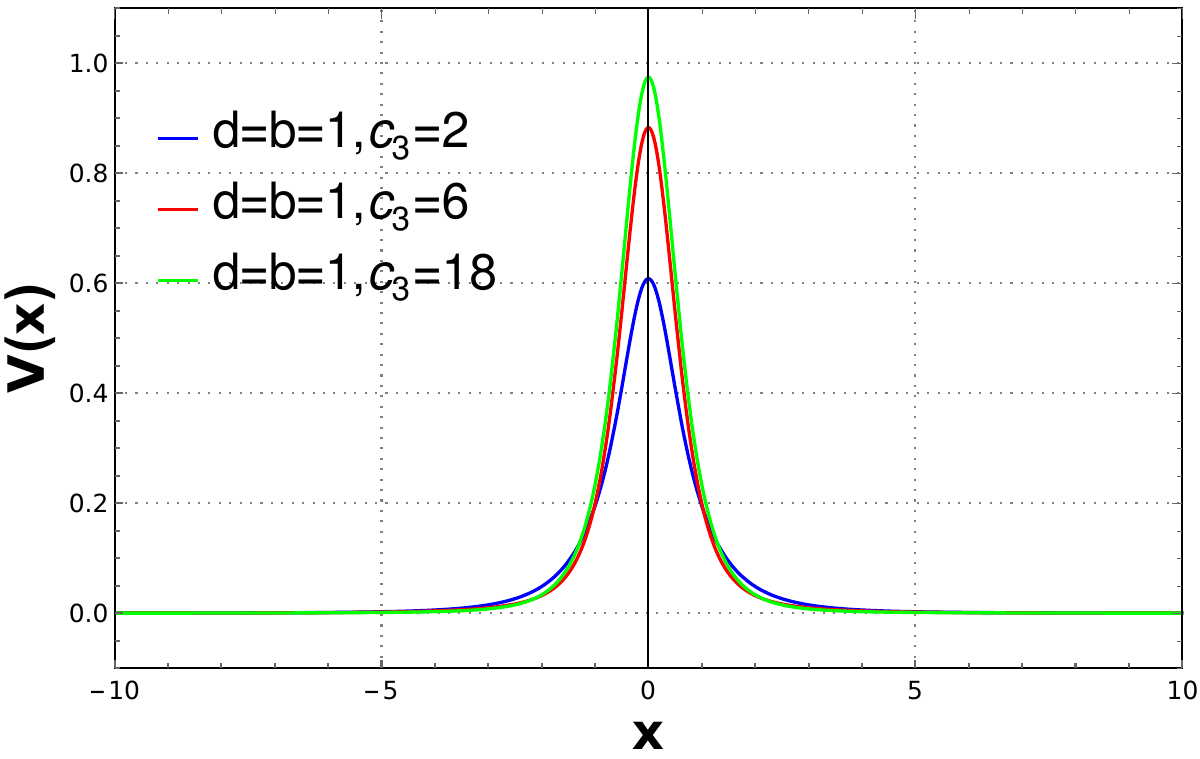}}}\qquad
    \subfigure[]{\label{POTEMOD23}
	{\includegraphics[width=0.3\linewidth]{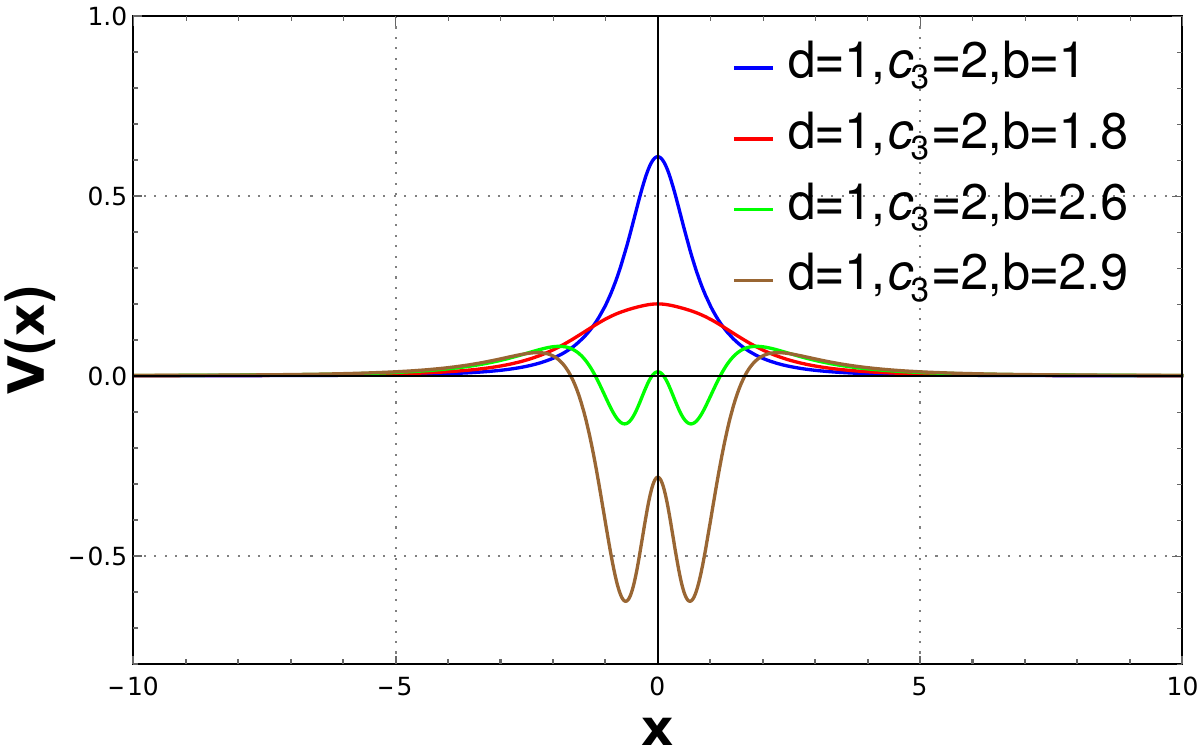}}}} \\
    \mbox{
	\subfigure[]{\label{LAGRAMOD21}
	{\includegraphics[width=0.3\linewidth]{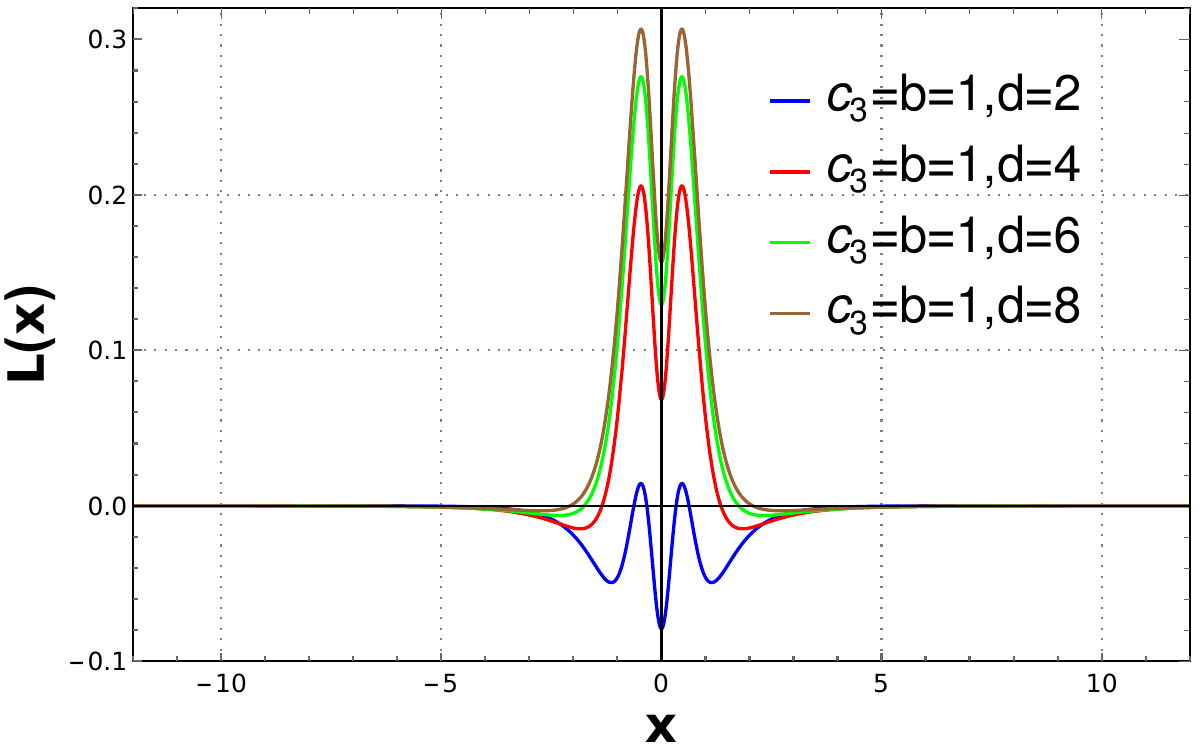}}}\qquad
	\subfigure[]{\label{LAGRAMOD22}
	{\includegraphics[width=0.3\linewidth]{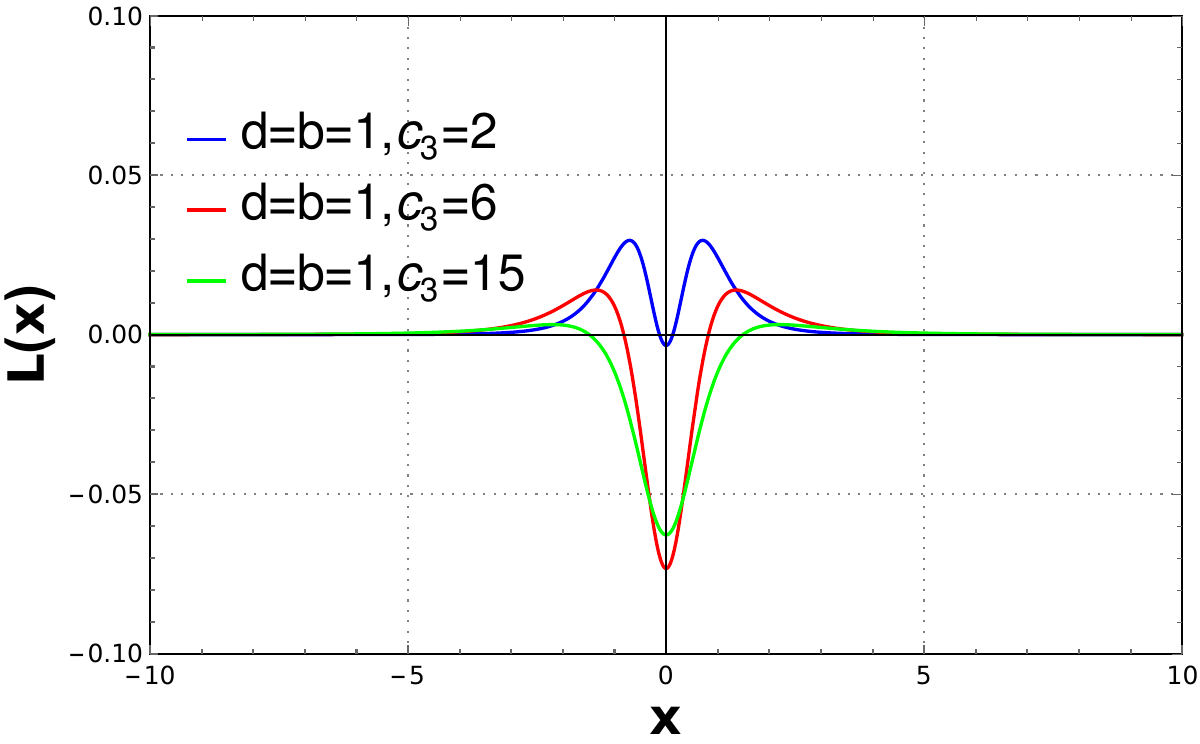}}}\qquad
    \subfigure[]{\label{LAGRAMOD23}
	{\includegraphics[width=0.3\linewidth]{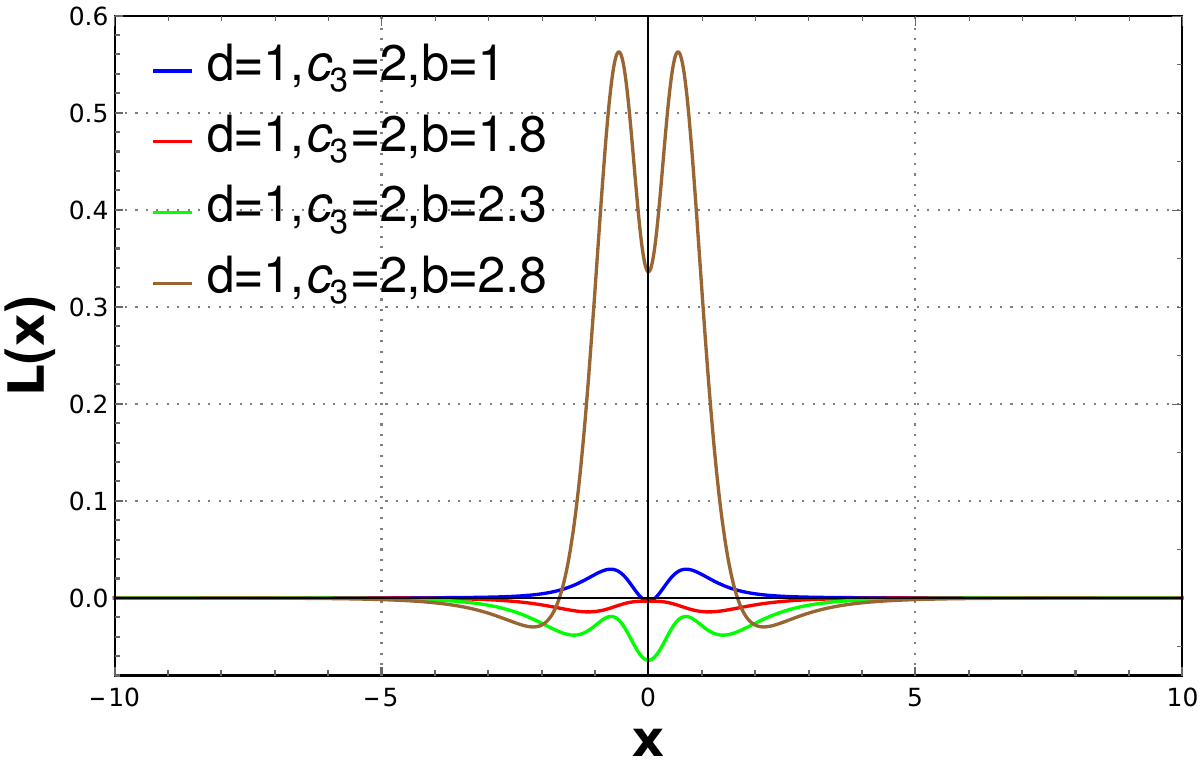}}}}
\caption{In the figures above the following values were set $n=1/2$ and $a=F_0=1$. In panels a), b), c) we have the variation of the parameters of the scalar field as a function of the radial coordinate. In panels d), e), f) we have the variation of the potential parameters as a function of the radial coordinate, the in panels g), h), i) we have the variation of the parameters of the Lagrangian function as a function of the radial coordinate.}
\label{FIG1MOD2}
\end{figure}

In Fig. \ref{FIG1MOD2} we have the graphical representation for the scalar field Eq. (\ref{CESCALARMOD2MAGN}), the potential Eq. (\ref{POTMOD2}) and Lagrangian Eq. (\ref{SECUNDMAGFUNCTIONMOD2}) functions with the variation of the model parameters related to the magnetically charged system.

As already mentioned at the beginning of the section, we fix the power $n=1/2$ for the k-essence field and thus, in Fig.(\ref{CAMPOMOD21}) we consider the parameters $c_3=b=1$ and then vary the parameter $d$. Thus, we can see that as we increase its value, the asymptotic behavior of the field attenuates when $x\to{+}\infty$ and increases when $x\to{-}\infty$, performing an oscillation at the origin of the coordinate system. Likewise, in Fig. (\ref{CAMPOMOD22}) we fix the parameters $d=b=1$ and then vary the coefficient $c_3$ where we can see that as we increase its value the scalar field grows in amplitude in the spatial asymptotics $x\to\pm{\infty}$. In Fig. (\ref{CAMPOMOD23}) we fixed the parameters $d=1$, $c_3=2$ and then we varied the parameter $b$ and we can see that as we increase it, the scalar field performs an oscillation at the origin of the coordinate system that tends to increase in amplitude as we increase the parameter values.

Regarding the behavior of the potential Eq. (\ref{POTMOD2}) we can observe in the Figs. (\ref{POTEMOD21}), (\ref{POTEMOD22}) and (\ref{POTEMOD23}) that when we vary the parameter $c_3$ and fix the others, it tends to behave like a barrier-type potential, while when we fix $c_3$ and vary the other parameters it tends to oscillate around the origin of the coordinate system. Depending on certain parameter values, the minimum points of the potential tend to be shifted towards the negative, which may indicate the possibility of the existence of normal or quasi-normal modes when subjected to perturbations. Finally, in Figs. (\ref{LAGRAMOD21}), (\ref{LAGRAMOD22}) and (\ref{LAGRAMOD23}) we have a representation of the Lagrangian function as a function of the radial coordinate with the variation of some parameters of the model.

\begin{figure}
    \centering
    \subfigure[]{\includegraphics[width=0.47\linewidth]{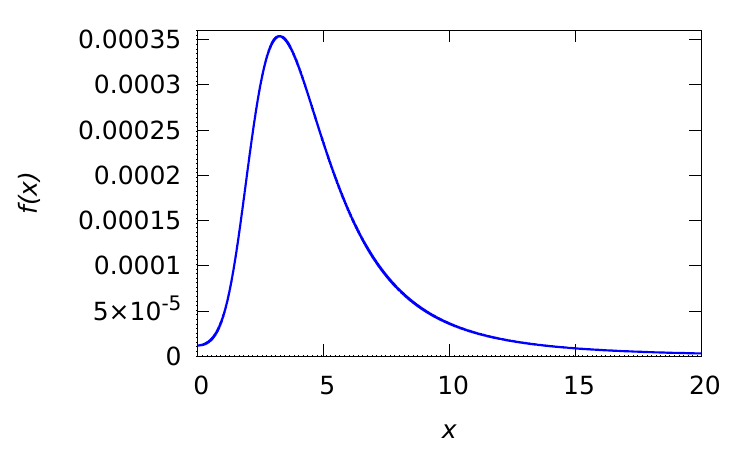}}
    \subfigure[]{\includegraphics[width=0.47\linewidth]{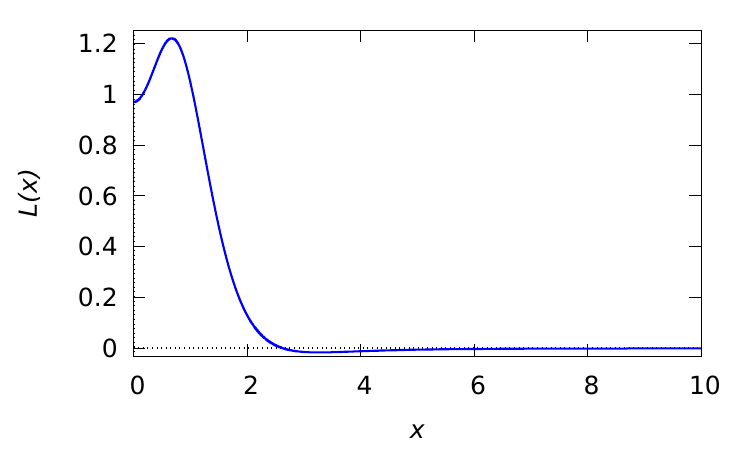}}
    \subfigure[]{\includegraphics[width=0.47\linewidth]{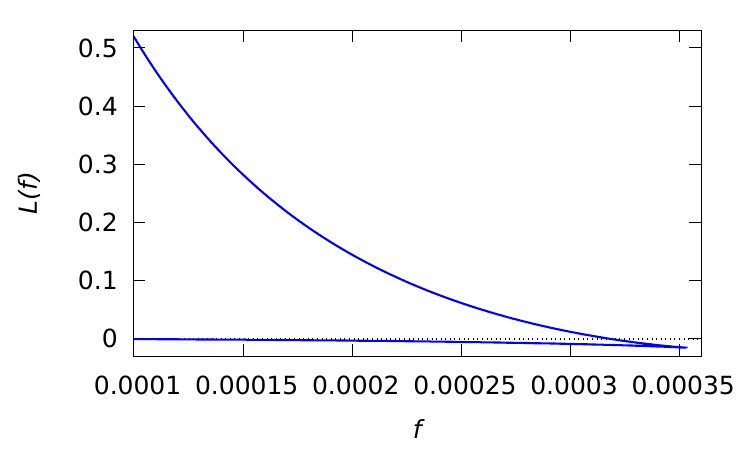}}
    \caption{In the figures above the following values were set $q_m=d=1$, $c_3=3$, and $b=4$. In a) we have the scalar $f(x)$, in b) the Lagrangian in terms of the radial coordinate $x$, and in c) the Lagrangian $L(f)$.}
    \label{fig:Lmag_model2}
\end{figure}

For this model, the electromagnetic scalar is given by
\begin{equation}
    f(x)=\frac{q_m^2 e^{-\frac{2 b^2}{c_3+x^2}}}{2 \left(d^2+x^2\right)^2}.
\end{equation}
Even for a magnetically charged source, it is not possible to invert the function $f(x)$. This becomes clearer when examining Fig. \ref{fig:Lmag_model2}. As we can see, the function $f(x)$ has a maximum value. However, whenever $f(x)$ exhibits a maximum or minimum, $L(f)$ becomes multivalued due to the inversion required to obtain $x(f)$. Therefore, it is not possible to write the Lagrangian $L(f)$ analytically for this model, even in the magnetically charged case. Depending on the values of the chosen parameters, the number of maxima and minima in $f(x)$ can increase, making the multivalued behavior of $L(f)$ even more pronounced. Consequently, the Lagrangian $L(f)$ may exhibit different behaviors in different regions of spacetime.

\subsection{Eletric case} \label{MOD2ELECTRO1}

\begin{figure}
    \centering
    \subfigure[]{\includegraphics[width=0.47\linewidth]{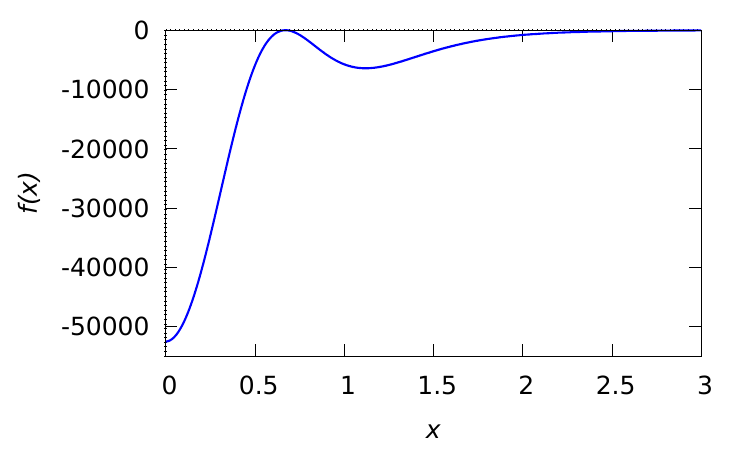}}
    \subfigure[]{\includegraphics[width=0.47\linewidth]{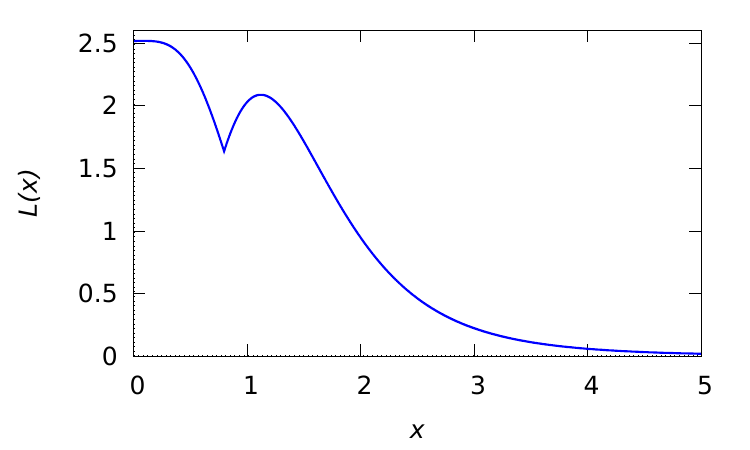}}
    \subfigure[]{\includegraphics[width=0.47\linewidth]{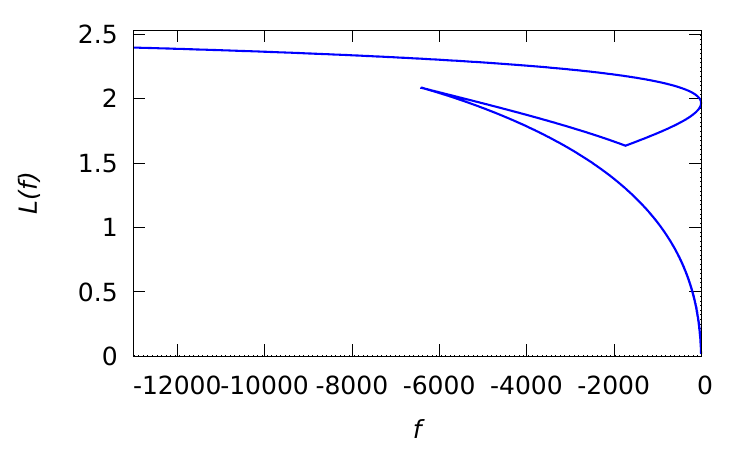}}
    \caption{In the figures above the following values were set $q_e=d=1$, $c_3=3$, and $b=4$. In a) we have the scalar $f(x)$, in b) the Lagrangian in terms of the radial coordinate $x$, and in c) the Lagrangian $L(f)$. All figures are considering the electric charged case.}
    \label{fig:Lele_model2}
\end{figure}

We will do here the electrically charged case where, as mentioned in the previous sections, the physical quantities related to the scalar field and the potential are the same as those obtained in the previous section (\ref{MOD2MG1}). To do this, we will first consider Eqs. (\ref{20}) and (\ref{21}) and then use the area function for the wormhole, as well as the scalar field Eq. (\ref{CESCALARMOD2MAGN}) and the potential Eq. (\ref{POTMOD2}). In the following, we present the electromagnetic functions for the system in which the matter source is electrically charged
\begin{equation}\label{ELETROFUNC1MOD2}
    L_f(x)= -\frac{q_e^2 \left(c_3+x^2\right){}^4 e^{-\frac{b^2}{c_3+x^2}}}{2 \left(d^2+x^2\right)I_1},
\end{equation}in which
\begin{eqnarray}\label{ELETROFUNC1MOD2PLUS}
 I_1&=& -\left(c_3+x^2\right){}^4 + e^{\frac{b^2}{c_3+x^2}} \left(2 b^4 x^2 \left(d^2+x^2\right)+b^2 \left(3 d^2 x^4-x^6\right)+x^8\right) \\\nonumber 
 &+&c_3 e^{\frac{b^2}{c_3+x^2}} \left(c_3 \left(-b^2 \left(d^2+5 x^2\right)+4 c_3 x^2+c_3^2+6 x^4\right)+2 b^2 x^2 \left(d^2-3 x^2\right)+4 x^6\right).
\end{eqnarray}
\begin{eqnarray} \nonumber\label{ELETROFUNC2MOD2}
    L(x)&=&\frac{b^2 \left(\frac{b^4 d^8 x^2}{\left(c_3+x^2\right){}^5}+12 d^2 (d-b) (b+d) \left(\log \left(d^2+x^2\right)-\log \left(c_3+x^2\right)\right)+12 c_3^2 \left(\log \left(c_3+x^2\right)-\log \left(d^2+x^2\right)\right)\right)}{\left(d^2-c_3\right){}^4} \\\nonumber
    &+&\frac{b^2 d^4 \left(c_3 \left(b^4 \left(d^4-20 d^2 x^2\right)+10 b^2 \left(3 d^4 x^2-19 d^2 x^4+30 x^6\right)+50 d^4 x^4\right)-30 x^6 (b-d) (b+d) \left(d^2-2 x^2\right)\right)}{5 \left(d^2-c_3\right){}^4 \left(c_3+x^2\right){}^5}  \\\nonumber 
    &+&\frac{2 b^2 c_3 d^2 \left(c_3 d^2 \left(-2 b^4 d^2+10 x^4 \left(33 b^2-19 d^2\right)+5 x^2 \left(3 b^4-17 b^2 d^2+d^4\right)\right)+30 x^6 \left(x^2 (d-b) (b+d)-5 d^4\right)\right)}{5 \left(d^2-c_3\right){}^4 \left(c_3+x^2\right){}^5} \\\nonumber
    &+&\frac{2 b^2 c_3^2 d^2 \left(c_3 \left(3 b^4 d^2-5 b^2 d^4+15 x^4 \left(4 d^2-17 b^2\right)-10 x^2 \left(b^4-24 b^2 d^2+5 d^4\right)-5 d^6\right)+30 \left(4 d^2 x^6+x^8\right)\right)}{5 \left(d^2-c_3\right){}^4 \left(c_3+x^2\right){}^5} \\
    &+&\frac{b^2 \left(-270 b^2 c_3^2 d^2 x^6+30 b^2 d^8 x^4+c_3^4 \left(b^4 \left(5 x^2-4 d^2\right)+60 b^2 d^4+40 d^6\right)-60 c_3^3 x^6 \left(x^2-5 d^2\right)\right)}{5 \left(d^2-c_3\right){}^4 \left(c_3+x^2\right){}^5} \\\nonumber 
    &+&\frac{b^2 c_3^4 \left(10 x^2 \left(b^2 \left(x^2-33 d^2\right)-24 d^4+70 d^2 x^2-27 x^4\right)+c_3 \left(b^4-30 b^2 d^2-180 d^4\right)\right)}{5 \left(d^2-c_3\right){}^4 \left(c_3+x^2\right){}^5}\\\nonumber 
    &-&\frac{2 b^2 c_3^5 \left(c_3 \left(2 b^2+13 c_3-28 d^2+41 x^2\right)+x^2 \left(b^2-74 d^2+49 x^2\right)\right)}{\left(d^2-c_3\right){}^4 \left(c_3+x^2\right){}^5},
\end{eqnarray}
while the electromagnetic scalar is given by
\begin{equation}
    f(x)=-\frac{2I_1^2}{q_e^2\left(c_3+x^2\right)^8}.
\end{equation}
From Fig. \ref{fig:Lele_model2}, we observe that, for the same parameter values, the functions associated with the electromagnetic sector are more complex in the electrically charged case than in the magnetically charged one. The function $f(x)$ displays both a maximum and a minimum, which by itself implies at least three distinct behaviors for $L(f)$. Moreover, the behavior of $L(x)$ shows another change in the electromagnetic Lagrangian. As a result, the plot of $L(f)$ clearly reveals four different behaviors. Therefore, even with the same parameter values, the Lagrangian exhibits an even more pronounced multivalued behavior in the electrically charged case compared to the magnetically charged case.

\section{THIRD MODEL}\label{sec5}
As a third model, we will consider the line element Eq. (\ref{line_x}) with $A(x)=1$ and $\Sigma^2(x)=b^2+\frac{x^{2m+2}}{d^{2m}+x^{2m}}$, where $m$ is a positive integer number. This area function is inspired by the black-bounce model presented in \cite{MR1CQG}. The main motivation for introducing this model is to minimize the violation of the energy conditions, specifically the null energy condition (NEC). Standard traversable wormholes require exotic matter that violates the NEC everywhere near the throat. This model arises as a possibility to confine this violation to a restricted region, ensuring that the NEC is satisfied at least in some regions of the spacetime, thus making the geometry physically more realistic. The line element that describes this model is written as
\begin{equation}\label{line3}
    ds^2=dt^2-dx^2-\left(b^2+\frac{x^{2m+2}}{d^{2m}+x^{2m}}\right)d\Omega^2.
\end{equation}
This spacetime is symmetric under the transformation $x \to -x$ and has a throat at $x=0$. The presence of the throat can be confirmed by verifying that
\begin{equation}
    \left.\frac{d^{2m+1}\Sigma^2}{dx^{2m+1}}\right|_{x=0}=0, \quad \mbox{and} \quad \left.\frac{d^{2m+2}\Sigma^2}{dx^{2m+2}}\right|_{x=0}>0.
\end{equation}

\begin{figure}[htb!]
\centering  
	\mbox{
	\subfigure[]{\label{Sigma3b}
	{\includegraphics[width=0.45\linewidth]{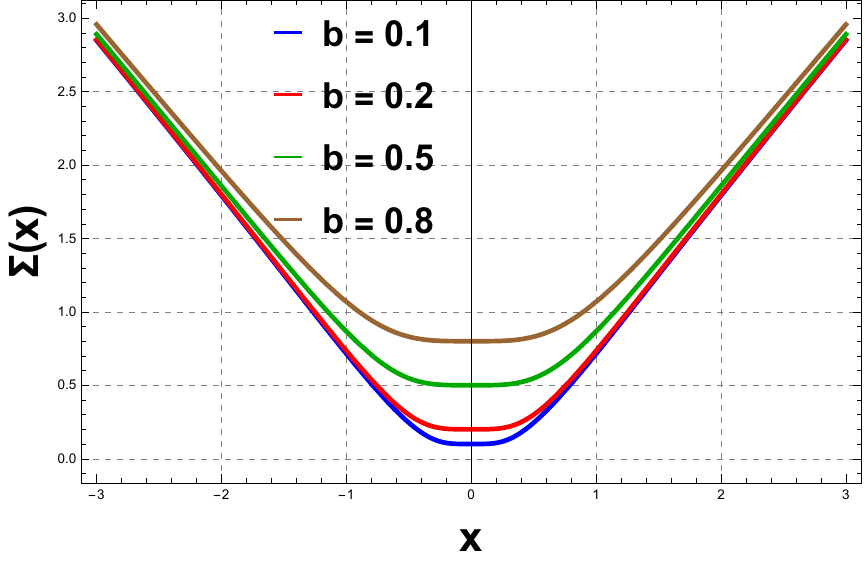}}}
    \subfigure[]{\label{Sigma3d}
	{\includegraphics[width=0.45\linewidth]{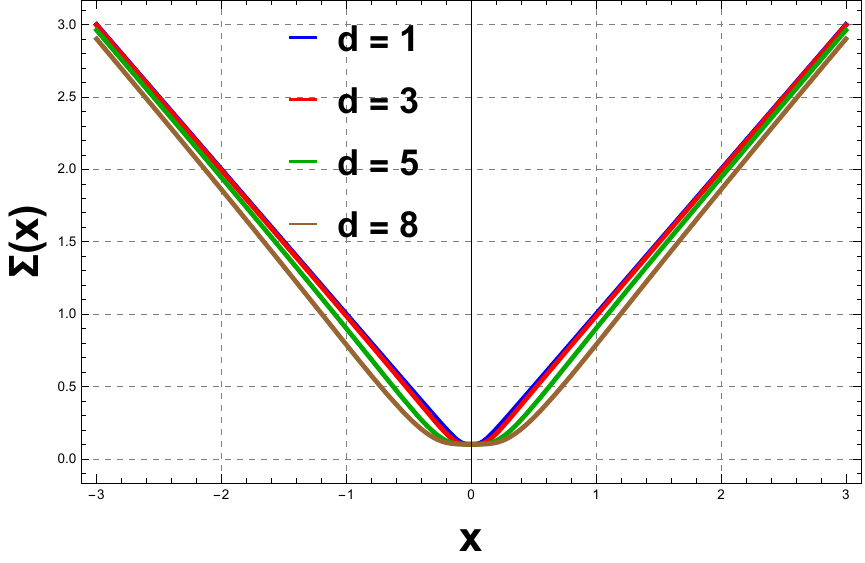}}}}
\caption{Behavior of $\Sigma$ of the spacetime \eqref{line3} as a function of the coordinate $x$. In panel~(a), we fix $d=1$ and vary the values of $b$. In panel~(b) we fix $b=0.1$ and vary the value of $d$.}
\label{FIGSigmaMOD3}
\end{figure}

In Fig.~\ref{FIGSigmaMOD3}, we show the behavior of $\Sigma$ as the parameters of the solution are varied. From panel~\ref{Sigma3b}, we see that the shape of the wormhole is mainly modified through changes in the throat radius, rather than in the throat position, when the parameter $b$ is varied. In panel~\ref{Sigma3d}, we observe that the effect produced by varying the parameter $d$ is relatively small, since it modifies neither the size nor the position of the wormhole throat. In addition, neither of these parameters changes the number of throats present in the geometry.

For both electrically and magnetically charged system configurations, we consider $\eta=-1$, $F_0=m=1$, and we fix the k-essence field strength at $n=1/2$.

\subsection{Magnetic case}\label{MOD3MG1}

For this case, already explained in Section (\ref{sec2A}), we consider the magnetically charged system. Therefore, for the generalized area function mentioned above and considering in equation (\ref{14}) the power $n=1/2$ for the field of k-essence, we then have the expression for the scalar field:
\begin{eqnarray}\label{CAMPOESCALARMOD3}
    \phi(x)&=& \frac{2 x \left(b^2+2 x^2\right)}{b^2 \left(d^2+x^2\right)+x^4}+\frac{2 \left(4 b^2-7 d^2\right) \tan ^{-1}\left(\frac{x}{d}\right)}{d^3}-\frac{6 x}{d^2+x^2} \\\nonumber 
    &+& \frac{4 \left(2 i b^3-2 b^2 \sqrt{4 d^2-b^2}+3 d^2 \sqrt{4 d^2-b^2}-6 i b d^2\right) \tan ^{-1}\left(\frac{\sqrt{2} x}{\sqrt{b \left(b-i \sqrt{4 d^2-b^2}\right)}}\right)}{d^2 \sqrt{8 d^2-2 b^2} \sqrt{b \left(b-i \sqrt{4 d^2-b^2}\right)}}\\\nonumber
    &+&\frac{4 \left(-2 i b^3-2 b^2 \sqrt{4 d^2-b^2}+3 d^2 \sqrt{4 d^2-b^2}+6 i b d^2\right) \tan ^{-1}\left(\frac{\sqrt{2} x}{\sqrt{b \left(b+i \sqrt{4 d^2-b^2}\right)}}\right)}{d^2 \sqrt{8 d^2-2 b^2} \sqrt{b \left(b+i \sqrt{4 d^2-b^2}\right)}}.
\end{eqnarray}

Taking into account Eq. (\ref{16}) and making use of the analyzed area function, along with the expression for the scalar field Eq. (\ref{CAMPOESCALARMOD3}), we then define the expression for the scalar potential:
\begin{eqnarray}\label{POTMOD3}
&&V(x)= \frac{2 \left(8 d^2-5 b^2\right)}{d^2 \left(d^2+x^2\right)}-\frac{b^2 x^2 \left(b^2-4 d^2\right)}{\left(b^2 \left(d^2+x^2\right)+x^4\right)^2}+\frac{-2 b^4+b^2 \left(5 d^2-2 x^2\right)+2 d^2 x^2}{d^2 \left(b^2 \left(d^2+x^2\right)+x^4\right)}+\frac{3 d^2}{\left(d^2+x^2\right)^2} \\\nonumber
&&+\frac{3 \left(\frac{2 \left(2 b^6-11 b^4 d^2+14 b^2 d^4-2 d^6\right) \tan ^{-1}\left(\frac{b^2+2 x^2}{b \sqrt{4 d^2-b^2}}\right)}{b \sqrt{4 d^2-b^2}}-\left(2 b^4-7 b^2 d^2+4 d^4\right) \left(2 \log \left(d^2+x^2\right)-\log \left(b^2 \left(d^2+x^2\right)+x^4\right)\right)\right)}{d^6}.
\end{eqnarray}

It is worth noting that the values of the free parameters that were fixed for both the magnetically charged system and the electrical case were simply a choice because of simplicity.  Looking only at our area function $\Sigma(x)$, we can observe that for the choice of $m=1$ and $d=0$, we have the area function corresponding to the Simpson-Visser model \cite{matt}.
On the other hand, looking at the expressions referring to the scalar field Eq. (\ref{CAMPOESCALARMOD3}) and the potential Eq. (\ref{POTMOD3}), we can see that in the limit where the parameter $d\to{0}$ these quantities diverge, which leads to the idea that an EB type wormhole does not support NED as a source, which being possible only with phantom scalar fields. Additionally, we have the constraint that $4d^2-b^2>0$.

In a manner analogous to the procedure developed in the previous sections, we can use equations (\ref{15}) and (\ref{17}) to obtain the electromagnetic functions of this model:
\begin{eqnarray}\label{DERILMAGMOD3}
L_{f}(x)= \frac{2 \left(d^6-3 d^4 x^2\right) \left(b^2 \left(d^2+x^2\right)+x^4\right)}{\left(d^2+x^2\right)^4 q_m^2}, 
\end{eqnarray}
\begin{eqnarray}\label{LMAGMOD3} 
L(x)&=& \frac{2 \left(\frac{8 b^2-13 d^2}{d^2+x^2}+\frac{4 b^4+b^2 \left(4 x^2-9 d^2\right)+d^4-5 d^2 x^2}{b^2 \left(d^2+x^2\right)+x^4}-\frac{2 d^4}{\left(d^2+x^2\right)^2}\right)}{d^2}\\\nonumber
&+& \frac{6 \left(2 b^4-7 b^2 d^2+4 d^4\right) \left(2 \log \left(d^2+x^2\right)-\log \left(b^2 \left(d^2+x^2\right)+x^4\right)\right)}{d^6}  \\\nonumber
&+& \frac{12 \left(-2 b^6+11 b^4 d^2-14 b^2 d^4+2 d^6\right) \tan ^{-1}\left(\frac{b^2+2 x^2}{b \sqrt{4 d^2-b^2}}\right)}{b d^6 \sqrt{4 d^2-b^2}},
\end{eqnarray} where $q_m$ represents the magnetic charge.

\begin{figure}[htb!]
\centering  
	\mbox{
	\subfigure[]{\label{CAMPOMOD31}
	{\includegraphics[width=0.45\linewidth]{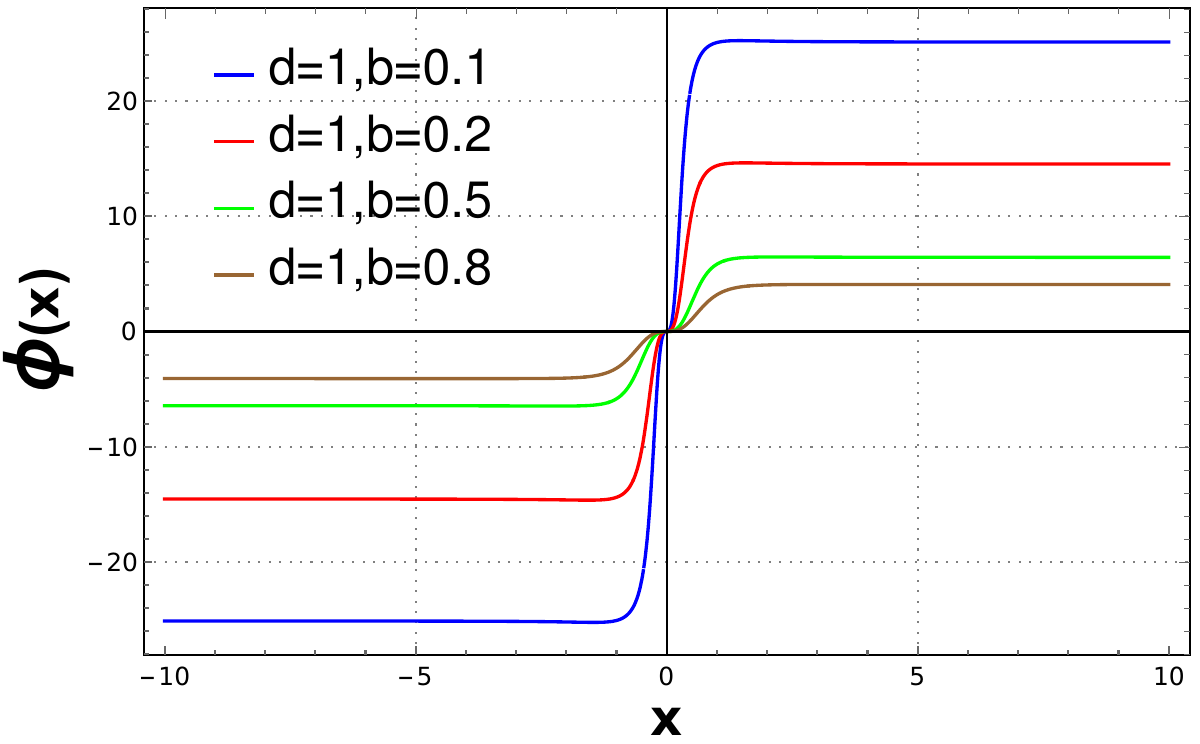}}}\qquad
	\subfigure[]{\label{CAMPOMOD32}
	{\includegraphics[width=0.45\linewidth]{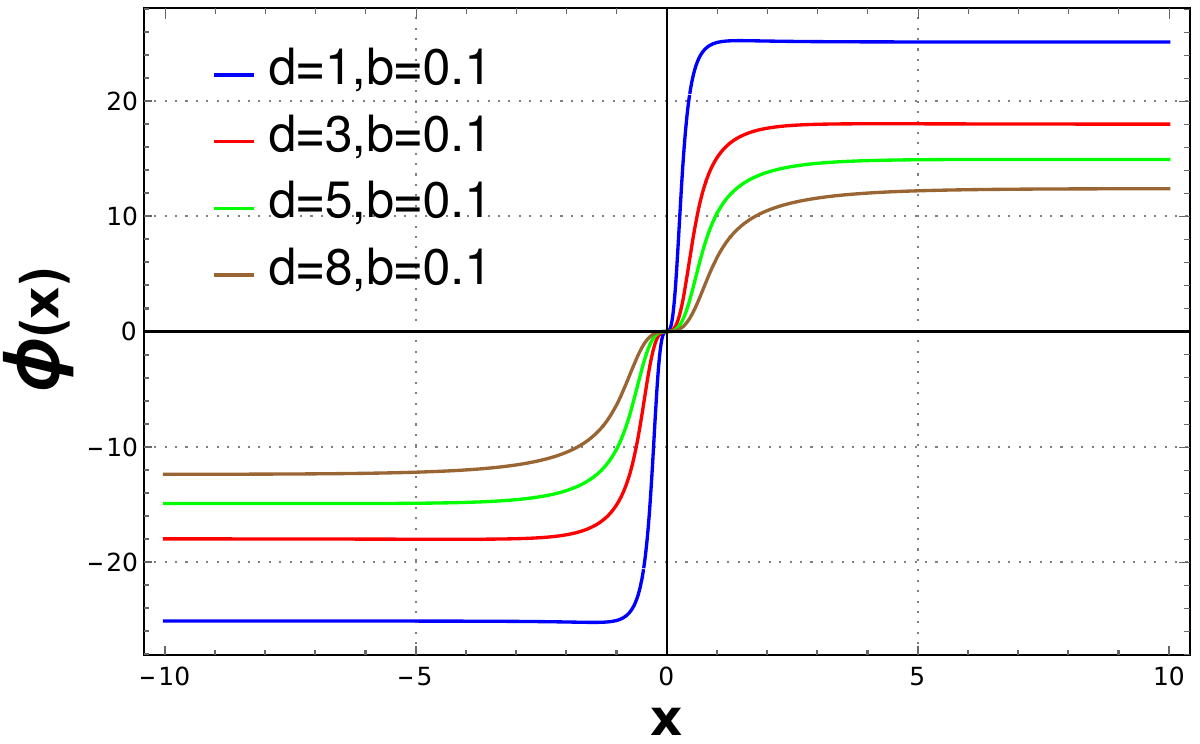}}}} \\
    \mbox{
	\subfigure[]{\label{POTEMOD31}
	{\includegraphics[width=0.45\linewidth]{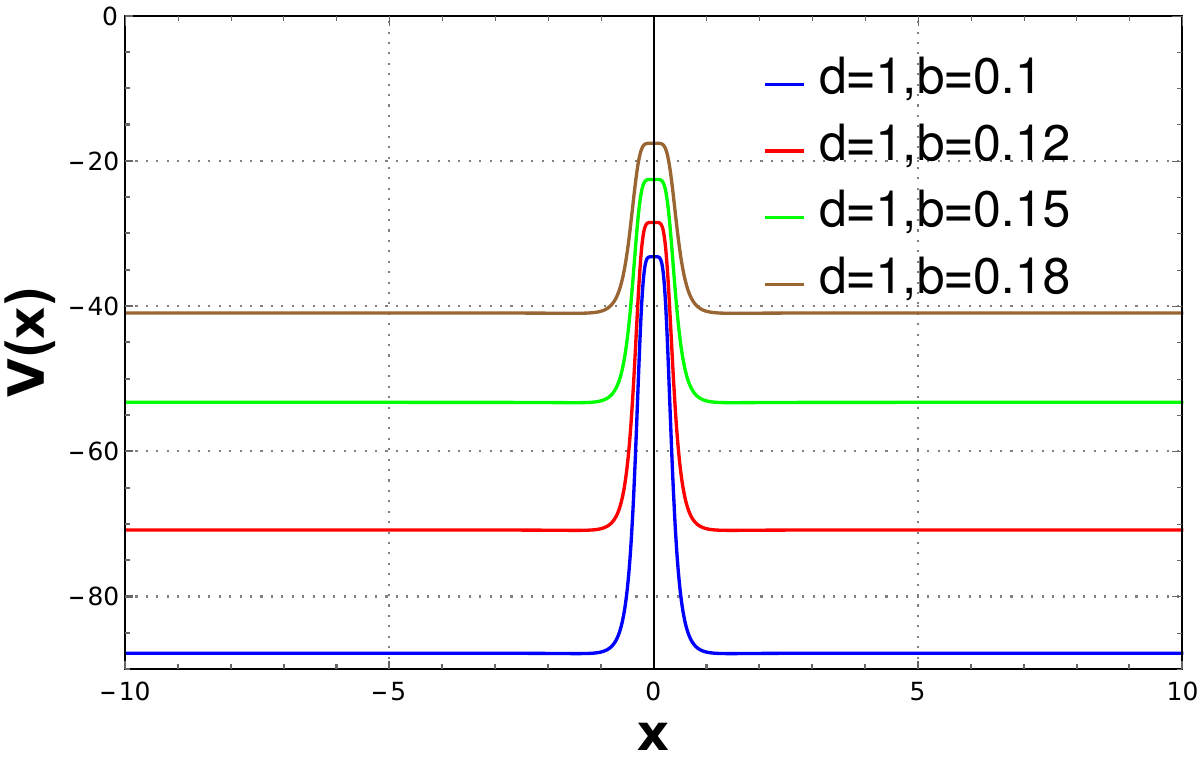}}}\qquad
	\subfigure[]{\label{POTEMOD32}
	{\includegraphics[width=0.45\linewidth]{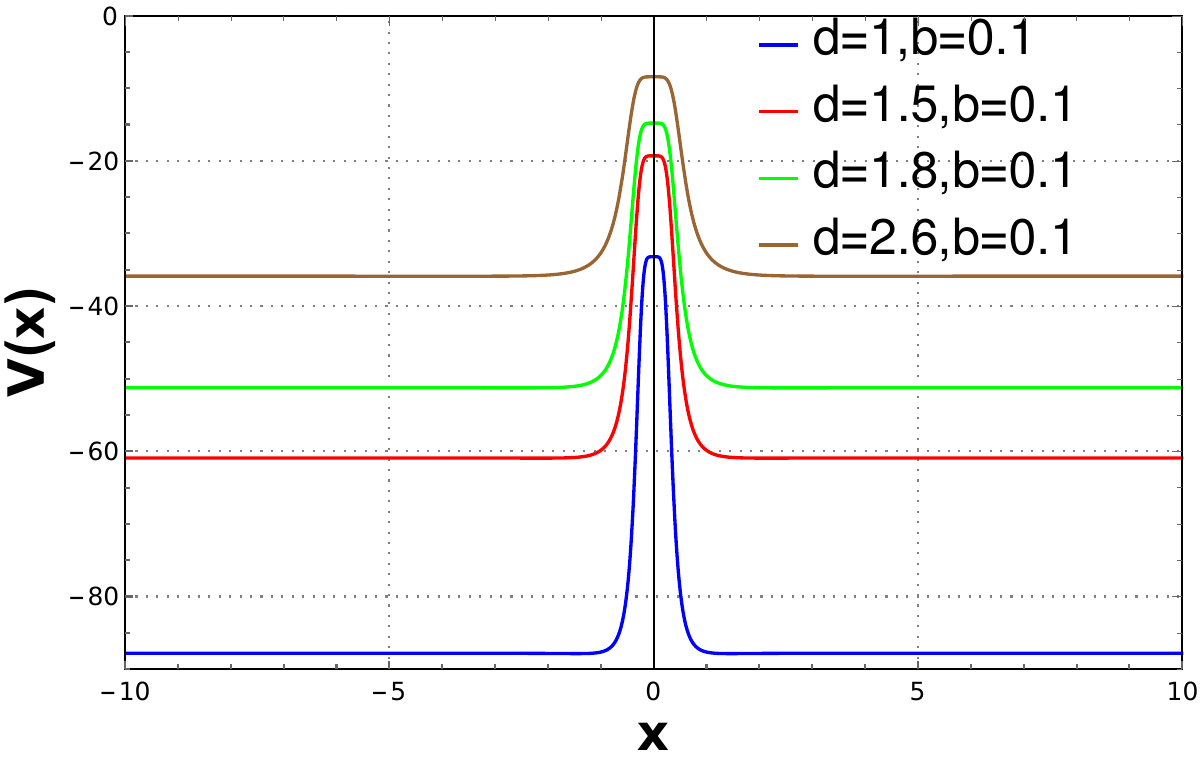}}}} \\
    \mbox{
	\subfigure[]{\label{LAGRAMOD31}
	{\includegraphics[width=0.45\linewidth]{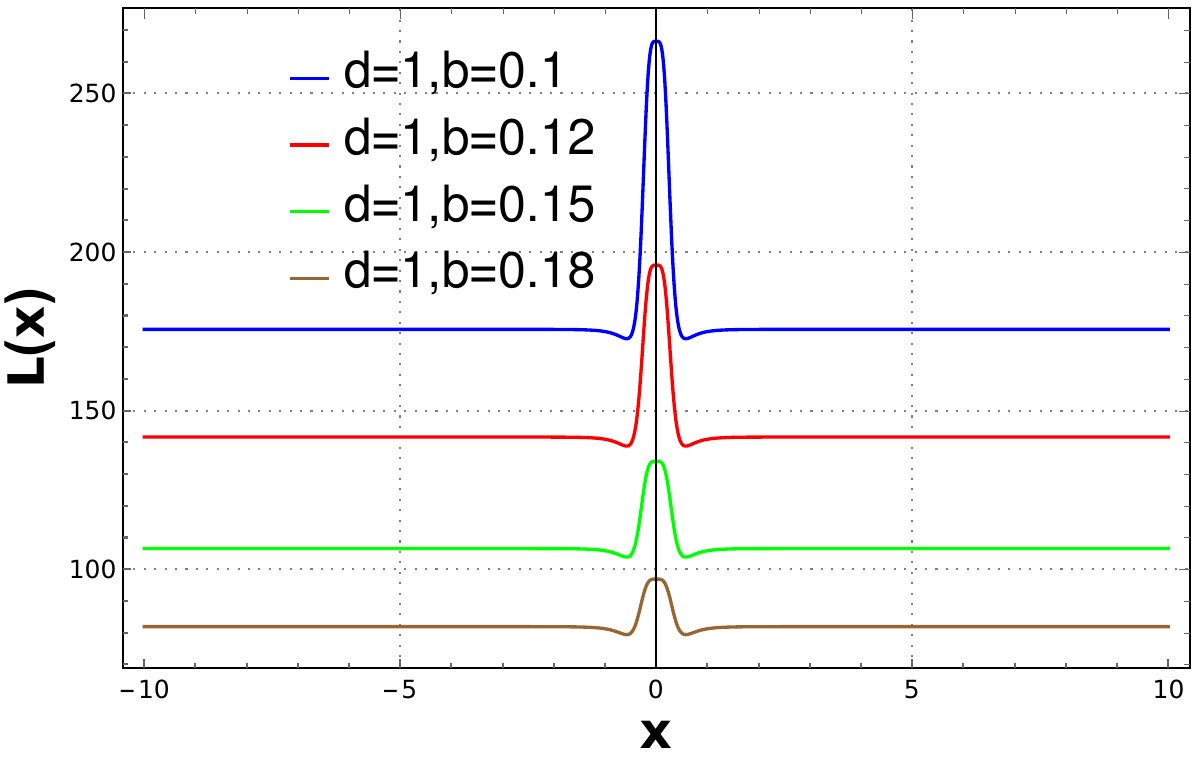}}}\qquad
	\subfigure[]{\label{LAGRAMOD32}
	{\includegraphics[width=0.45\linewidth]{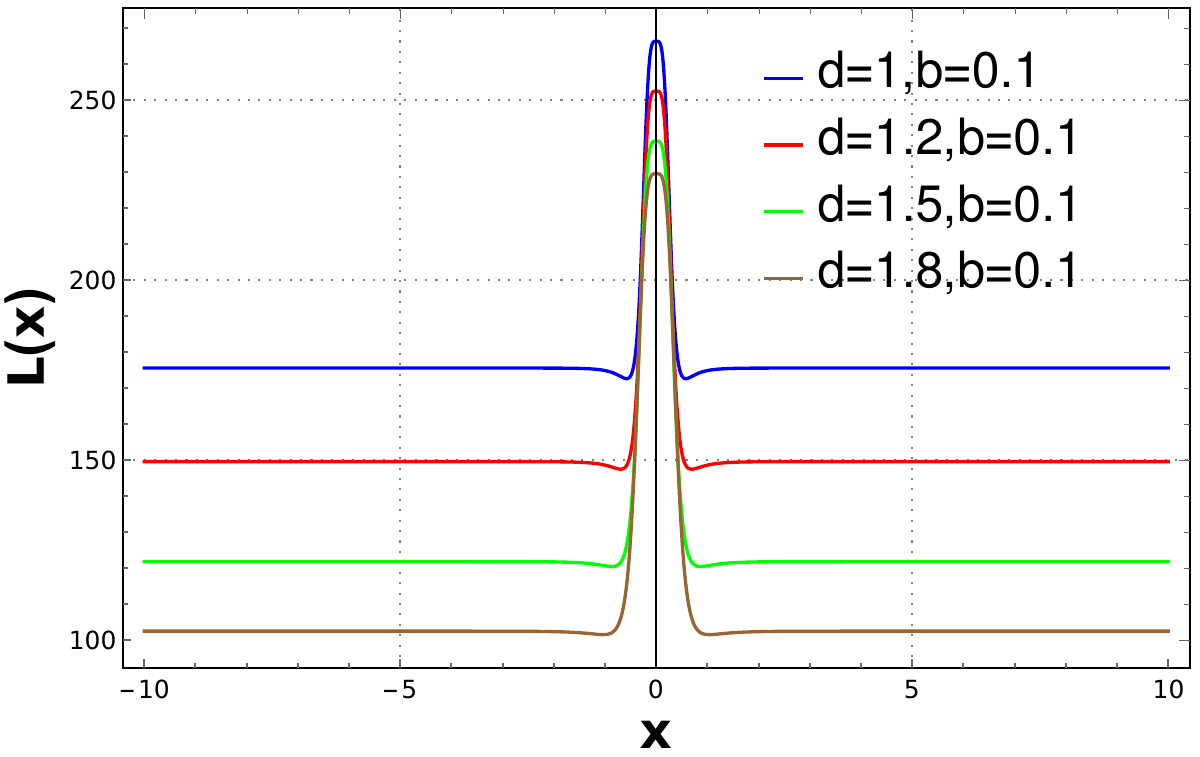}}}}
\caption{In the figures above, the following values were defined: $n=1/2$, $m=q_m=F_0=1$ and $\eta=-1$. In panels a) and b), we have the variation of the scalar field parameters as a function of the radial coordinate. In panels c) and d), we have the variation of the potential parameters as a function of the radial coordinate, and in panels e) and f), we have the variation of the Lagrangian function parameters as a function of the radial coordinate.}
\label{FIG1MOD3}
\end{figure}

In Fig. (\ref{FIG1MOD3}), we have the graphical representation of the scalar field Eq. (\ref{CAMPOESCALARMOD3}), the potential Eq. (\ref{POTMOD3}) and the Lagrangian Eq. (\ref{LMAGMOD3}) referring to the magnetically charged system for the area function in question. We set the power of the k-essence field to $n=1/2$, and the parameters $F_0=m=q_m=1$ and $\eta=-1$. Thus, in figures (\ref{CAMPOMOD31}) and (\ref{CAMPOMOD32}), we have the behavior of the scalar field, where in the first figure we fix the parameter $d$ and then vary the parameter $b$, and in the second figure we do the opposite. Thus, we can see that in both cases the behavior is similar: when we increase the value of the parameter as $x\to{\pm}{\infty}$, the field tends to be damped in amplitude. On the other hand, in figures (\ref{POTEMOD31}) and (\ref{POTEMOD32}) we have the representation of the potential where the variation of the parameters serves to vertically shift the potential, which has a barrier-like behavior and behaves symmetrically in the asymptotic regions as $x\to{\pm\infty}$. Similarly, we observe the behavior of the Lagrangian function in figures (\ref{LAGRAMOD31}) and (\ref{LAGRAMOD32}). We can also highlight that everything indicates that even by varying the free parameters of the model, a minimum point for the potential is not obtained in a negative region, which may indicate that the scalar field is unstable when subjected to perturbations.

\subsection{Electric case} \label{MOD3ELECTRO1}

For the electrically charged case, we have that the scalar field Eq. (\ref{CAMPOESCALARMOD3}) and the potential Eq. (\ref{POTMOD3}) are the same as those established in the previous subsection (\ref{MOD3MG1}). Thus, considering the power of the k-essence field $n=1/2$ and the other parameters that have already been fixed above, we can combine equations (\ref{20}) and (\ref{21}) together with the area function to obtain the electromagnetic functions for this system. In this way, we have expressions relating to the source of electrically charged matter
\begin{equation}\label{DERILELETRICMODEL3}
    L_f(x)=\frac{q_e^2 \left(d^2+x^2\right)^4}{2 d^4 \left(d^2-3 x^2\right) \left(b^2 \left(d^2+x^2\right)+x^4\right)},
\end{equation}

\begin{eqnarray}\label{LELETRICMODEL3}
    L(x)&=& \frac{6 \left(\frac{4 b^2 d^4 \left(d^2+x^2\right)-8 d^8-6 d^6 x^2}{\left(d^2+x^2\right)^2}+\frac{\left(-4 b^6+22 b^4 d^2-28 b^2 d^4+4 d^6\right) \tan ^{-1}\left(\frac{b^2+2 x^2}{b \sqrt{4 d^2-b^2}}\right)}{b \sqrt{4 d^2-b^2}}\right)}{d^6} \\\nonumber
    &+& \frac{6 \left(2 b^4-7 b^2 d^2+4 d^4\right) \left(2 \log \left(d^2+x^2\right)-\log \left(b^2 \left(d^2+x^2\right)+x^4\right)\right)}{d^6},
\end{eqnarray} where $q_e$ represents the electric charge.

\begin{figure}[htb!]
\centering  
	\mbox{
	\subfigure[]{\label{LAGELETRICMOD31}
	{\includegraphics[width=0.3\linewidth]{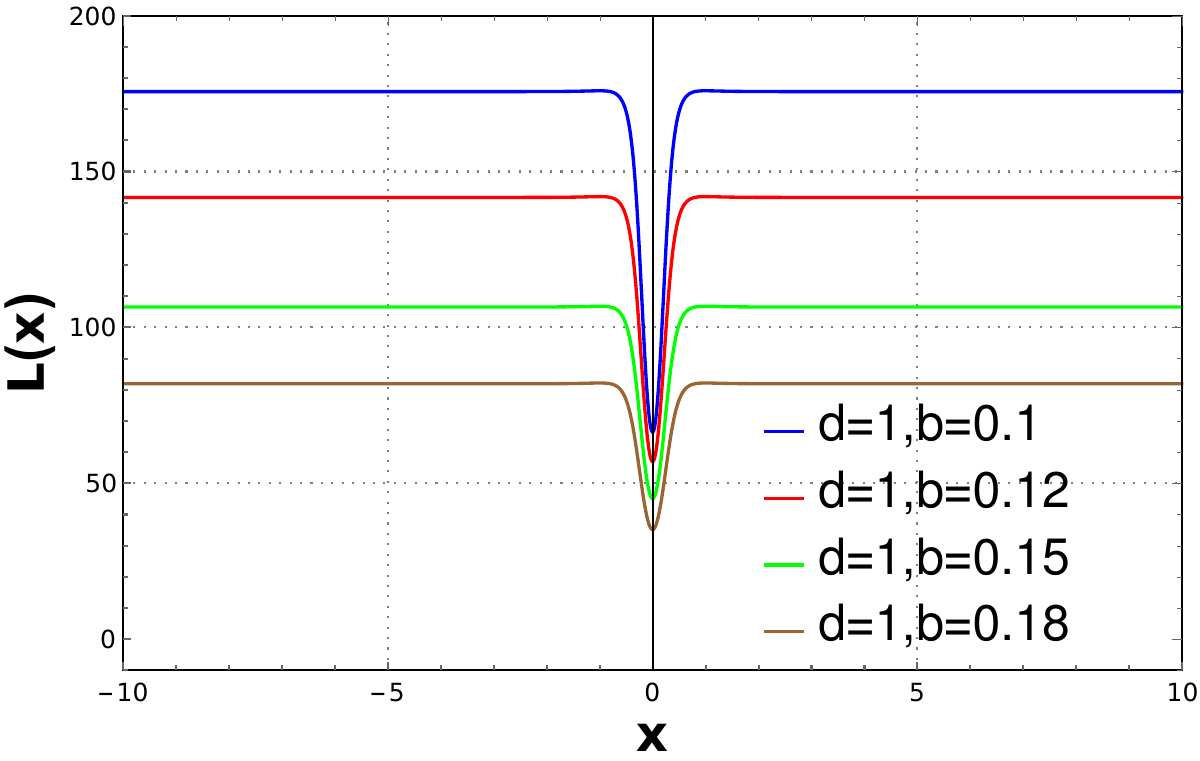}}}\qquad
	\subfigure[]{\label{LAGELETRICMOD32}
	{\includegraphics[width=0.3\linewidth]{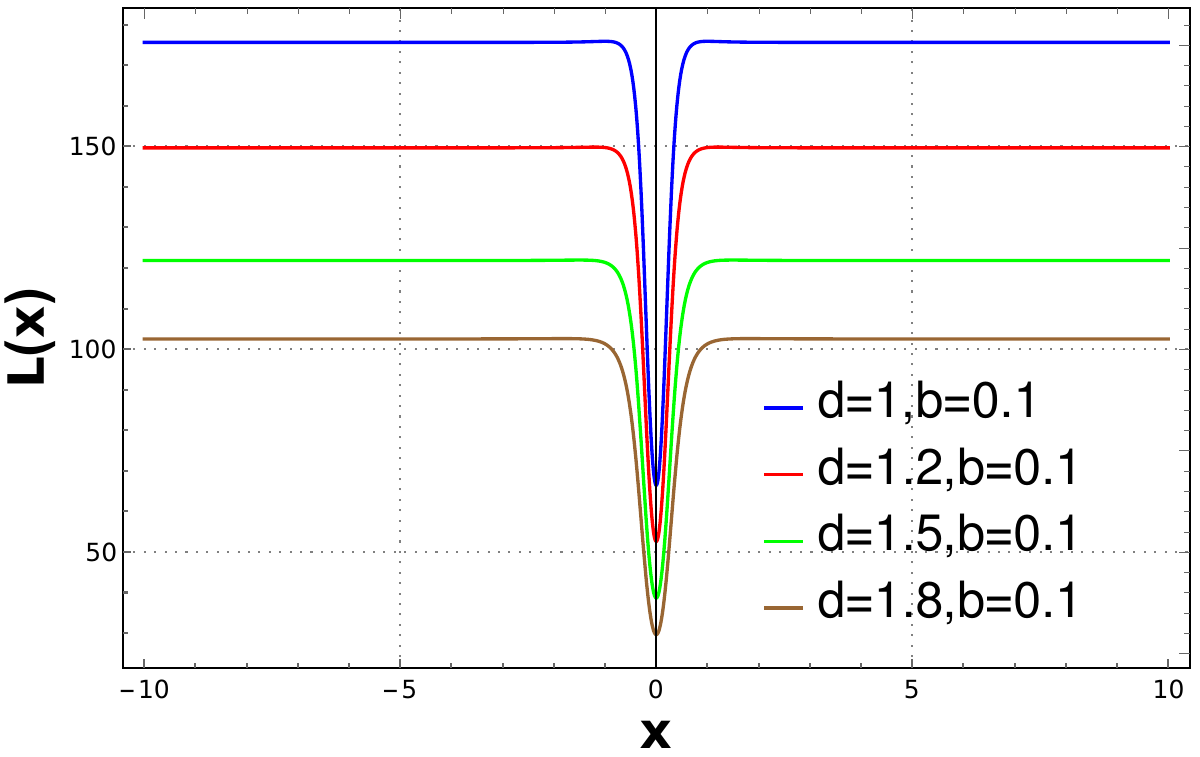}}}\qquad
	\subfigure[]{\label{LAGELETRICMOD33}
	{\includegraphics[width=0.3\linewidth]{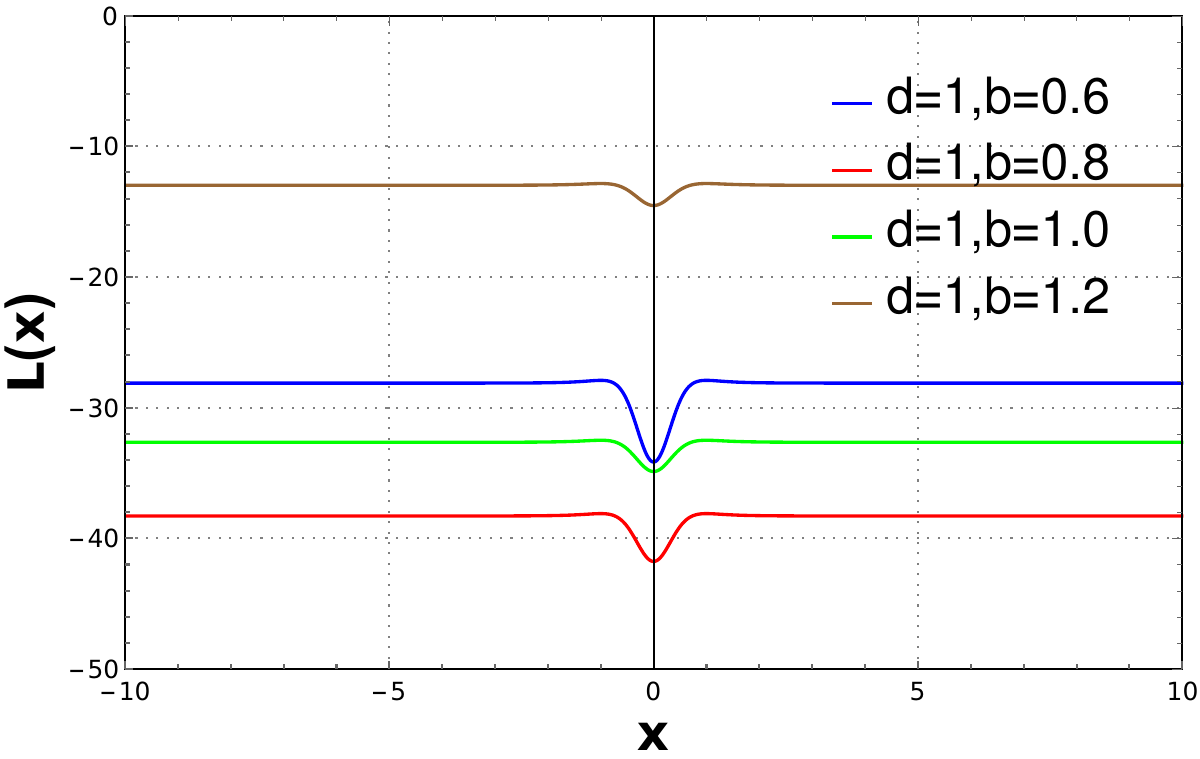}}}}
\caption{In the figures above, the following values were defined: $n=1/2$, $m=q_e=F_0=1$ and $\eta=-1$. In panels a), b) and c) we have the variation of the Lagrangian function for different parameters as a function of the radial coordinate.}
\label{FIG2MOD3}
\end{figure}

The scalar field and the potential for the electrically charged system, as in the previous models, are the same for the magnetically charged system. Therefore, what changes are only the expression of the Lagrangian Eq. (\ref{LELETRICMODEL3}) and its derivative Eq. (\ref{DERILELETRICMODEL3}). Thus, the graphical representation of the Lagrangian of this system is illustrated in Fig. \ref{FIG1MOD3}, where we fix the parameters similarly to the previous section (\ref{MOD3MG1}) and then vary the free parameters $b$ and $d$, whose main function is to vertically shift the curves as we vary the parameters, see Figs. (\ref{LAGELETRICMOD31}), (\ref{LAGELETRICMOD32}) and (\ref{LAGELETRICMOD33}).

\begin{figure}
    \centering
    \subfigure[]{\includegraphics[width=0.47\linewidth]{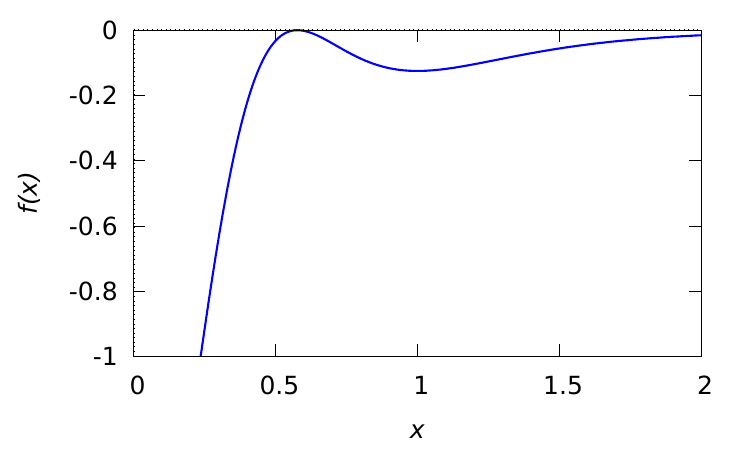}}
    \subfigure[]{\includegraphics[width=0.47\linewidth]{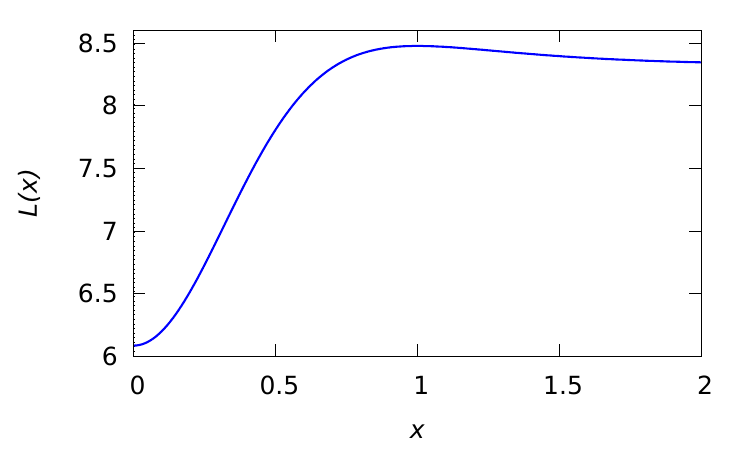}}
    \subfigure[]{\includegraphics[width=0.47\linewidth]{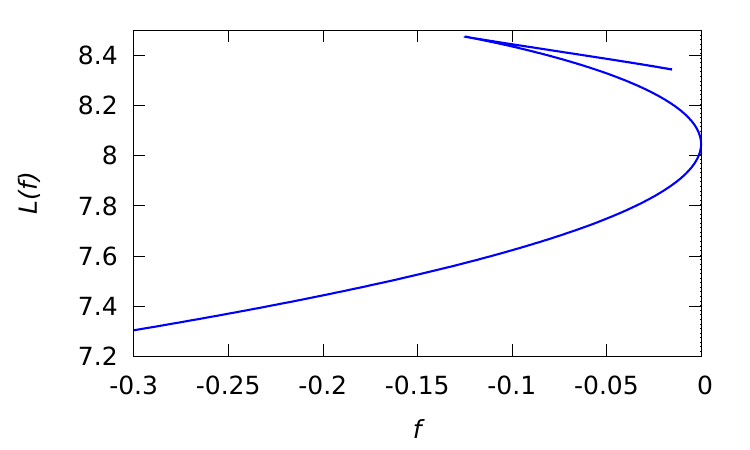}}
    \caption{In the figures above the following values were set $q_e=d=b=m=1$ and $n=1/2$. In a) we have the scalar $f(x)$, in b) the Lagrangian in terms of the radial coordinate $x$, and in c) the Lagrangian $L(f)$. All figures are considering the electric charged case.}
    \label{fig:Lele_model3}
\end{figure}

In addition to analyzing the electromagnetic functions by varying the solution parameters, we must also study their behavior to understand the form of $L(f)$. In Fig. \ref{fig:Lele_model3}, the behavior of $f(x)$ clarifies why an analytic expression for $L(f)$ is not feasible. The plot of $f(x)$ exhibits both maxima and minima, which prevents inversion to $x(f)$ and thus to $L(f)$. Since $f(x)$ shows one maximum and one minimum, the Lagrangian $L(f)$ must display at least three distinct branches: a smooth transition associated with $f(x)=0$ and a cusp-like transition when $f(x)\neq 0$.

\section{GENERAL ENERGY CONDITIONS}\label{sec6}
For static and spherically symmetric traversable wormholes in GR, it is a well-established result that the matter threading the throat must violate the NEC. The requirement of traversability in wormhole geometries implies the existence of a throat connecting two distant regions of spacetime. According to the Raychaudhuri equation, preventing the gravitational collapse of the throat requires the congruence of null geodesics to satisfy the flaring-out condition, which in general relativity implies a violation of NEC, namely $T_{\mu \nu} k^\mu k^\nu < 0$ for some null vector $k^\mu$ \cite{Hochberg:1997wp,Morris:1988cz,Visser:1995cc}. In this context, the matter source supporting the wormhole geometry is usually referred to as exotic matter, since it violates the standard energy conditions. In particular, this corresponds to a radial tension whose magnitude exceeds the energy density, a feature naturally provided by the k-essence and phantom scalar fields considered in this work. As a consequence, at least one of the remaining pointwise energy conditions is necessarily violated in a finite neighbourhood of the throat. To discuss this systematically for the metric (\ref{line_x}) and the sources defined in Eqs.~(\ref{4})–(\ref{5}), we write the usual energy conditions directly in terms of the energy density and the radial and tangential pressures. We denote by NEC, WEC, SEC and DEC the null, weak, strong and dominant energy conditions, respectively. The subscripts $1$, $2$ and $3$ indicate whether the corresponding combination involves the radial pressure, the tangential pressure, or both pressures (or only the energy density). With this notation, we have
\begin{eqnarray}\label{55}
NEC^{\phi,EM}_{1,2}&=&WEC^{\phi,EM}_{1,2}=SEC^{\phi,EM}_{1,2} \iff \rho^{\phi,EM} + p^{\phi,EM}_{1,2} \geq 0, \\\label{56}
SEC^{\phi,EM}_3 &\iff& \rho^{\phi,EM} + p^{\phi,EM}_{1} + 2p^{\phi,EM}_{2} \geq 0, \\\label{57}
DEC^{\phi,EM}_{1,2} &\iff&  \rho^{\phi,EM} + p^{\phi,EM}_{1,2} \geq 0  \qquad    \mbox{and} \qquad \rho^{\phi,EM} - p^{\phi,EM}_{1,2} \geq 0 , \\\label{58}
DEC^{\phi,EM}_3&=&WEC^{\phi,EM}_{3} \iff  \rho^{\phi,EM}  \geq 0 ,
\end{eqnarray}
where the superscripts $\phi$ and $EM$ label the contributions from the scalar field and the electromagnetic sector, respectively.

For the wormhole geometries considered here, the effective stress-energy tensor can be written in the anisotropic-fluid form
\begin{eqnarray}\label{59}
\tensor{T}{^\mu}_{\nu}= {\rm diag}\left[\rho^{\phi,EM},-p^{\phi,EM}_1,-p^{\phi,EM}_2,-p^{\phi,EM}_2\right],
\end{eqnarray}
with $\rho^{\phi,EM}$ the total energy density, and $p^{\phi,EM}_1$ and $p^{\phi,EM}_2$ denoting the radial and tangential pressures, respectively. Since traversable wormholes do not possess an event horizon and the coordinates remain static across the throat, the diagonal form (\ref{59}) is valid throughout the entire spacetime.

The fluid quantities to the electromagnetic stress-energy tensor, considering a magnetic charged source, are
\begin{equation}\label{60}
   \rho^{EM}= \frac{L(x)}{2}, \quad
     {p}^{EM}_1= -\frac{L(x)}{2}, \quad
     {p}^{EM}_2= -\frac{L(x)}{2} + \frac{q^2_m{L_f(x)}}{2\Sigma^4}.
\end{equation}
Although we consider a magnetically charged source, the energy conditions do not depend on the type of charge of the source.

To the scalar field, as $A(x)=1$, we have
\begin{equation}\label{61}
    \rho^\phi =-p_2^\phi= -\frac{{\Sigma}''}{n\Sigma} + V(x),\quad
    {p}^\phi_1 = \frac{{\Sigma}''}{\Sigma}\left(\frac{1}{n}-2\right) - V(x).
\end{equation}

The energy conditions for the scalar field field are given by
\begin{eqnarray}\label{62}
NEC^{\phi}_{1}&=& -\frac{2{\Sigma}''}{\Sigma} \geq 0, \quad
NEC^{\phi}_{2}= 0,  \\\label{63}
DEC^{\phi}_{1}&=&  \frac{2{\Sigma}''}{\Sigma}\left(1-\frac{1}{n}\right) + 2V(x) \geq 0,  \quad
DEC^{\phi}_{2}=  - \frac{2{\Sigma}''}{n\Sigma} + 2V(x) \geq 0,  \\\label{64}
DEC^{\phi}_{3}&=&  - \frac{{\Sigma}''}{n\Sigma} + V(x) \geq 0, \quad SEC^{\phi}_{3}= \frac{2{\Sigma}''}{\Sigma}\left(\frac{1}{n}-1\right)-2V(x) \geq 0.
\end{eqnarray}
Likewise, for the electromagnetic part
\begin{eqnarray}\label{65}
NEC^{EM}_{1}&=& 0, \quad
NEC^{EM}_{2}= \frac{2-(\Sigma^2)''}{2\Sigma^2} \geq 0,  \\\label{66}
DEC^{EM}_{1}&=& L(x) \geq 0,  \quad
DEC^{EM}_{2}=  L(x) - \frac{2-(\Sigma^2)''}{2\Sigma^2}\geq 0,  \\\label{67}
DEC^{EM}_{3}&=&  \frac{L(x)}{2} \geq 0,
\quad SEC^{EM}_{3}= - L(x) + \frac{2-(\Sigma^2)''}{\Sigma^2} \geq 0.
\end{eqnarray}
Thus, to check whether all energy conditions are satisfied, we would need to analyze twelve conditions for each solution, six for the scalar field and six for the electromagnetic field. However, as we can see, both $NEC_{2}^{\phi}$ and $NEC_{1}^{EM}$ are identically satisfied. Moreover, $DEC_{1}^{EM} = 2DEC_{3}^{EM}$, thereby reducing the total to nine expressions to be analyzed. 

If we wish to further simplify our analysis of the energy conditions, we can focus solely on the NEC, since, according to Eq.~(\ref{55}), it is included in all the others. Thus, if the NEC is violated, the others will be as well. Since $NEC_{2}^{\phi}$ and $NEC_{1}^{EM}$ are identically satisfied, we must examine the behavior of $NEC_{1}^{\phi}$ and $NEC_{2}^{EM}$ for the three models. For the GEB model, we always have $\Sigma''>0$, so $NEC_{1}^{\phi}$ is always violated. For the second and third models, depending on the choice of parameters, it is possible to have $\Sigma''<0$ in some regions, but not throughout the spacetime. Thus, even if the NEC can be satisfied in certain regions, the scalar field will always violate all energy conditions in at least some region of the spacetime.

To be more specific about the possibility of regions where the NEC is violated for each model, we must analyze the form of the expressions associated with the NEC in each case. As discussed previously, the quantity $-\frac{2\Sigma''}{\Sigma}$ is always negative for the GEB case. For the remaining models, the general expression becomes considerably more complicated. Nevertheless, we can investigate the asymptotic behavior in order to gain further insight. For the second wormhole model \eqref{line2}, we have
\begin{equation}
\begin{split}
    -\frac{2\Sigma''}{\Sigma}\approx\frac{2 \left(b^2 d^2-c_3^2\right)}{c_3^2 d^2}+O\left(x^2\right),\quad \mbox{if} \quad x\to 0, \\
    -\frac{2\Sigma''}{\Sigma}\approx -\frac{2 \left(b^2+d^2\right)}{x^4}+O\left(\frac{1}{x^5}\right), \quad \mbox{if} \quad x\to \infty.
\end{split}
\end{equation}
This result shows that, for the second model, this expression can become positive near $x\to0$, depending on the relation between the parameters, thus satisfying the inequality associated with $\mathrm{NEC}_{1}^{\phi}$. However, the expression is always negative in the asymptotic region $x\to\infty$. Therefore, independently of the choice of parameters, the NEC will necessarily be violated at least in distant regions.

If we perform the same expansions for the third model \eqref{line3}, we obtain that
\begin{equation}
\begin{split}
    -\frac{2\Sigma''}{\Sigma}\approx -\frac{2\left(1+m\right)\left(1+2m\right)}{mb^2d^2}x^{2m}+O\left(x^{2m+2}\right),\quad \mbox{if} \quad x\to 0, \\
    -\frac{2\Sigma''}{\Sigma}\approx -\frac{2 \left(b^2-d^2\right)}{x^4}+O\left(\frac{1}{x^6}\right), \quad \mbox{if} \quad m=1 \quad \mbox{and} \quad  x\to \infty, \\
    -\frac{2\Sigma''}{\Sigma}\approx  -\frac{2 m\left(m-1\right)}{x^2}+O\left(\frac{1}{x^4}\right), \quad \mbox{if} \quad m=1 \quad \mbox{and} \quad x\to \infty.
\end{split}
\end{equation}
In this way, we can see that in the region $x\to0$, where the throat of this wormhole is located, the inequality associated with $\mathrm{NEC}_{1}^{\phi}$ is not satisfied. For asymptotic regions, however, the result depends on the values of $m$. If $m=1$, the inequality is satisfied provided that $b<d$, whereas for $m>1$ it is not satisfied in this region.

For all three models, there will always be regions where $2-(\Sigma^2)''<0$, making $NEC_{2}^{EM}$ negative somewhere. Thus, although $NEC_{1}^{EM}=0$, the NEC is also violated for the electromagnetic field. Therefore, both the scalar and electromagnetic fields violate all energy conditions, since they violate the NEC at least in some regions.

\section{Linear scalar stability analysis}

In this section, we shall study the stability of the three models considered in this work through quasi-normal modes and the time-domain evolution of a test scalar field. The equation of motion governing the dynamics of the test scalar field is the massless Klein–Gordon equation. When the effective potential possesses only a single maximum, we may employ the well-known sixth-order WKB method to determine the quasi-normal mode (QNM) oscillation frequencies numerically and approximately. However, when the effective potential exhibits two or more maxima, this method must either be generalized or replaced by another approach for evaluating scalar linear stability. Therefore, in cases where the potential possesses two maxima, for instance, which corresponds to the majority of the more general cases of our models, we shall investigate scalar linear stability through the time-domain evolution method of the test scalar field.

We consider a non--massive scalar field $\Psi$ on a static, spherically symmetric background metric
\eqref{line_x} with $A(x)=1$. The field obeys the Klein--Gordon equation 
\begin{eqnarray}
&&\Box \Psi = \frac{1}{\sqrt{-g}} \partial_\mu \left( \sqrt{-g} g^{\mu\nu} \partial_\nu \Psi \right)=0,\\
&&\Psi(t,x,\theta,\phi)=\frac{\psi(x,t)}{\Sigma(x)} Y_{l,m}(\theta,\phi) = e^{-i\omega t} \frac{\psi(x)}{\Sigma(x)} Y_{l,m}(\theta,\phi).
\end{eqnarray}
The radial equation reduces to
\begin{eqnarray}
\frac{d^2\psi}{dx_*^2} + \left[ \omega^2 - V_{eff}(x_*) \right] \psi = 0,\label{eqSchrodinger}
\end{eqnarray}
where $dx_* = dx\sqrt{-g_{11}/g_{00}}=dx$ is the tortoise coordinate, and $V_{eff}(x)$ is the effective potential given by
\begin{eqnarray}
V_{eff}(x)=\frac{l(l+1)}{\Sigma^2(x)}+\frac{1}{2\Sigma(x)}\frac{d}{dx}\left[\frac{1}{\Sigma(x)}\frac{d\Sigma^2(x)}{dx}\right]\,.    
\end{eqnarray}

The WKB method for QNMs approximates the complex frequencies $\omega_n$ via the Bohr--Sommerfeld type condition. Up to sixth order \cite{IyerWill1987,Konoplya2003}:
\begin{eqnarray}
\frac{iQ_0}{\sqrt{2Q_0''}} - \sum_{k=2}^6 \Lambda_k = n + \frac12, \quad n=0,1,2,\dots
\end{eqnarray}
Here $Q(\omega,x_*) = \omega^2 - V_{eff}(x_*)$, $Q_0 = Q(\omega,x_0)$, with $x_0$ the peak of $V_{eff}(x_*)$. The $\Lambda_k$ are corrections involving derivatives of $V$ up to order $2k$ evaluated at $x_0$. For sixth order, $\Lambda_2,\dots,\Lambda_6$ are given explicitly in standard references; they yield highly accurate $\omega_n$ for $\ell \geq n$.

Define null coordinates $u = t - x_*$, $v = t + x_*$. The wave equation $\frac{\partial^2 \psi(x,t)}{\partial t^2} - \frac{\partial^2 \psi(x,t)}{\partial x_*^2} + V_{eff}(x)\psi(x,t) = 0$ becomes
\begin{eqnarray}
\frac{\partial^2 \psi(u,v)}{\partial u \partial v} + \frac14 V_{eff}(u,v) \psi(u,v) = 0.\label{TD1}
\end{eqnarray}
This is the characteristic form, ideal for integration along null rays.

On a double--null grid $(u_i, v_j)$ with spacing $\Delta = \Delta u = \Delta v$, the finite--difference scheme \cite{Gundlach1994} is:
\begin{eqnarray}
\psi_{i+1,j+1} = \psi_{i+1,j} + \psi_{i,j+1} - \psi_{i,j} - \frac{\Delta^2}{8} V_{eff}(x_*)\big( \psi_{i+1,j} + \psi_{i,j+1} \big) + \mathcal{O}(\Delta^4),
\end{eqnarray}
where $x_* = (v_j - u_i)/2$. The evolution proceeds from initial data on $u=u_0$ and $v=v_0$.

The initial conditions are fixed by
\begin{eqnarray}
\psi(0,v)=\exp\left[-\frac{(v-v_c)^2}{2\sigma^2}\right]\;,\;\psi(u,0)=0\,.    
\end{eqnarray}
where \(v_c\) is associated with the center of the Gaussian wave packet and \(\sigma\) denotes its width.

Given the late--time signal $\psi(t)$ at a fixed spatial point, we assume a superposition of QNMs:
\begin{eqnarray}
\psi(t) \approx \sum_{p=1}^{P} C_p e^{-i\omega_p t}.\label{TD2}
\end{eqnarray}
The Prony method fits this model by solving a linear system for the frequencies. Steps:
\begin{enumerate}
    \item Sample $\psi(t)$ at equally spaced times $t_k = t_0 + k h$.
    \item Construct the Hankel matrix from the data and solve for the coefficients of the characteristic polynomial whose roots are $z_p = e^{-i\omega_p h}$.
    \item Extract $\omega_p = i\frac{\ln z_p}{h}$.
\end{enumerate}
This yields the dominant QNM frequencies from the time--domain waveform, confirming stability when all $\operatorname{Im}(\omega_p) < 0$.

The effective potentials for each model are
\begin{equation}
V_{\rm eff}^{(I)}(x)
=
-\frac{(m-1)x^{2m-2}}
{\left(a^{m}+x^{m}\right)^2}
+
\frac{(m-1)x^{m-2}}
{a^{m}+x^{m}}+
\ell(\ell+1)
\left(a^{m}+x^{m}\right)^{-2/m}.\label{eqVeff1}
\end{equation}
\begin{eqnarray}
&&V_{\rm eff}^{(II)}(x)
=
\frac{
e^{-b^2/(c_3+x^2)}
\,\ell(\ell+1)
}
{d^2+x^2}+
\frac{
c_3^2(c_3-bd)(c_3+bd)
}
{
(d^2+x^2)(c_3+x^2)^4
}+
\frac{
\left[c_3^2(4c_3-5b^2)
+2b^2(b^2+c_3)d^2\right]x^2
}
{
(d^2+x^2)(c_3+x^2)^4
}
\nonumber\\
&&
+
\frac{
(2b^4+6c_3^2+3b^2(d^2-2c_3))x^4
}
{
(d^2+x^2)(c_3+x^2)^4
}
+
\frac{
(4c_3-b^2)x^6+x^8
}
{
(d^2+x^2)(c_3+x^2)^4
}-
\frac{
x^2\left[(c_3+x^2)^2-b^2(d^2+x^2)\right]^2
}
{
(d^2+x^2)^2(c_3+x^2)^4
}.\label{eqVeff2}
\end{eqnarray}
\begin{equation}
V_{\rm eff}^{(III)}(x)
=
\frac{
x^{2+2m}\,{\cal P}_1(x)
+
b^2\left(d^{2m}+x^{2m}\right){\cal P}_2(x)
}
{
\left(d^{2m}+x^{2m}\right)^2
\left[
x^{2+2m}
+b^2\left(d^{2m}+x^{2m}\right)
\right]^2
}.
\end{equation}
\begin{equation}
{\cal P}_1(x)
=
d^{6m}\ell(\ell+1)+d^{4m}
\Bigl[
3\ell(\ell+1)+m+m^2
\Bigr]
x^{2m}
+d^{2m}
\Bigl[
3\ell(\ell+1)+m-2m^2
\Bigr]
x^{4m}
+\ell(\ell+1)x^{6m},
\end{equation}
\begin{eqnarray}
{\cal P}_2(x)
&=&d^{6m}\ell(\ell+1)+d^{4m}
\Bigl[3\ell(\ell+1)+(1+m)(1+2m)
\Bigr]
x^{2m}
\nonumber\\
&&+d^{2m}\Bigl[2+3\ell(\ell+1)+(3-2m)m\Bigr]x^{4m}+\left(1+\ell+\ell^2\right)x^{6m}.\label{eqVeff3}
\end{eqnarray}

\begin{figure}[htb]
    \centering

    \subfigure[]{
        \includegraphics[width=0.47\linewidth]{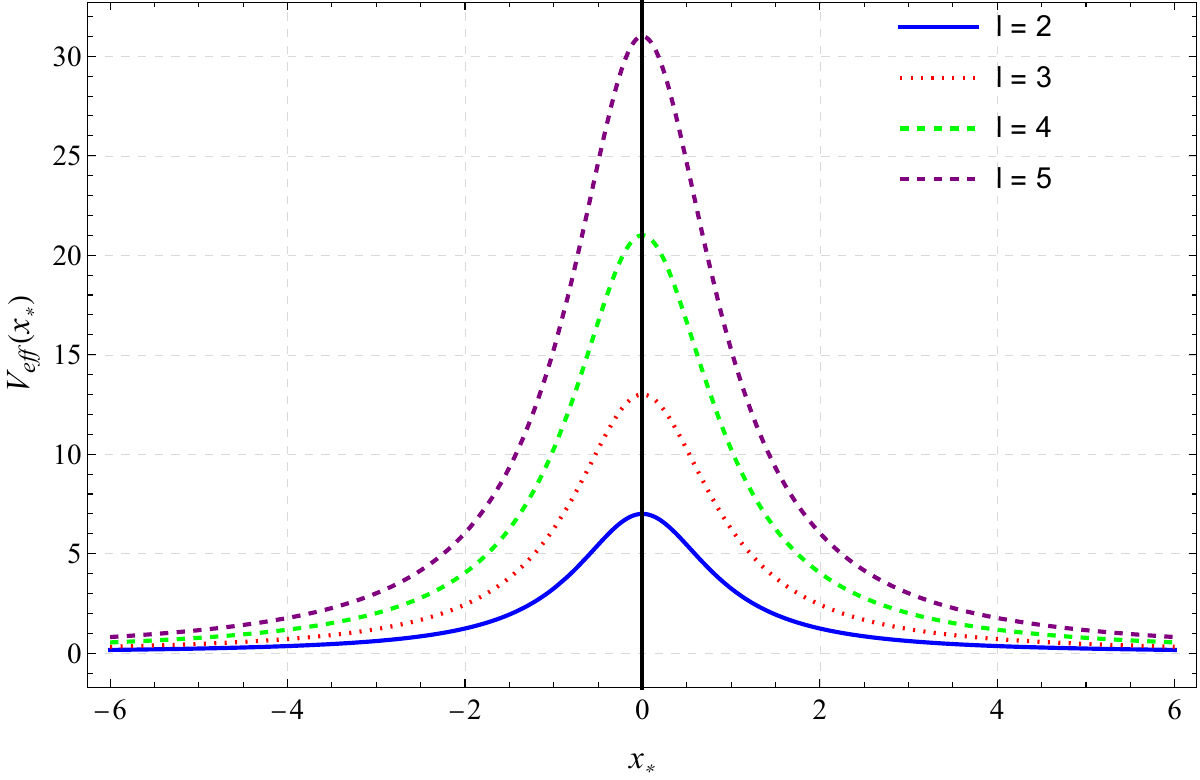}
    }
    \hfill
    \subfigure[]{
        \includegraphics[width=0.47\linewidth]{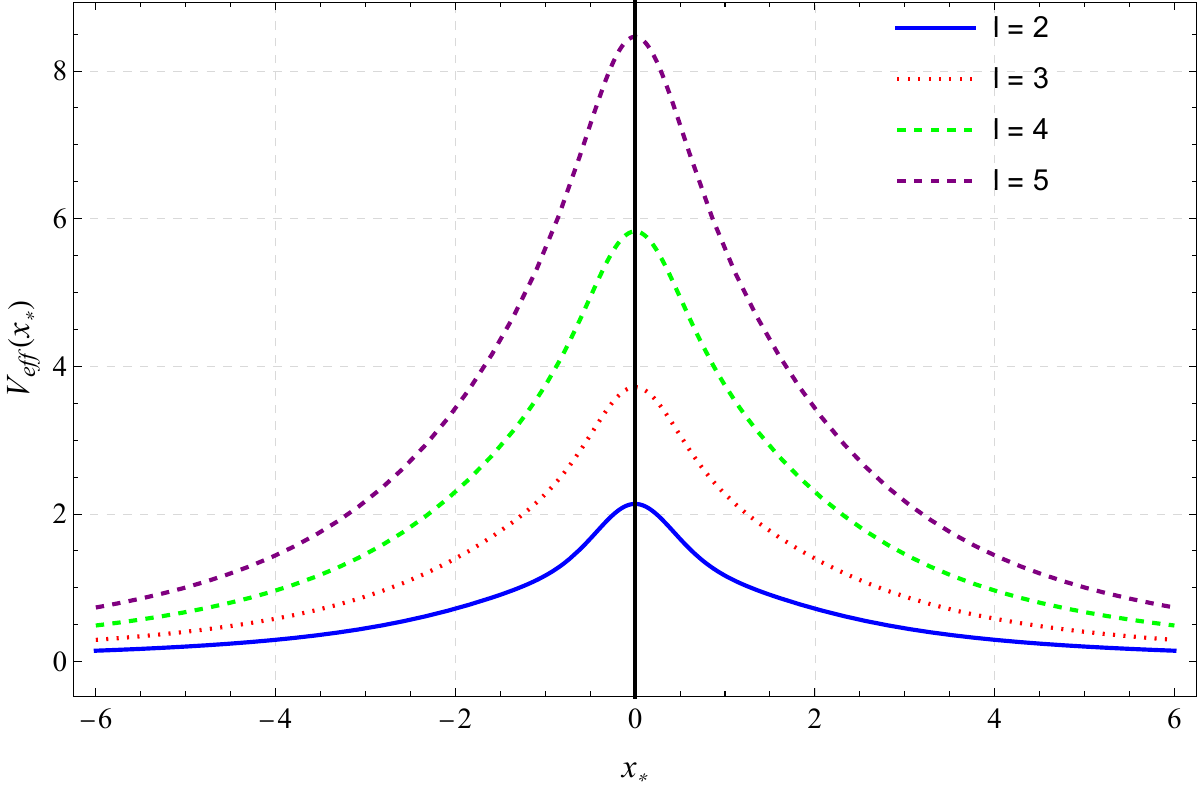}
    }

    \vspace{0.3cm}

    \subfigure[]{
        \includegraphics[width=0.47\linewidth]{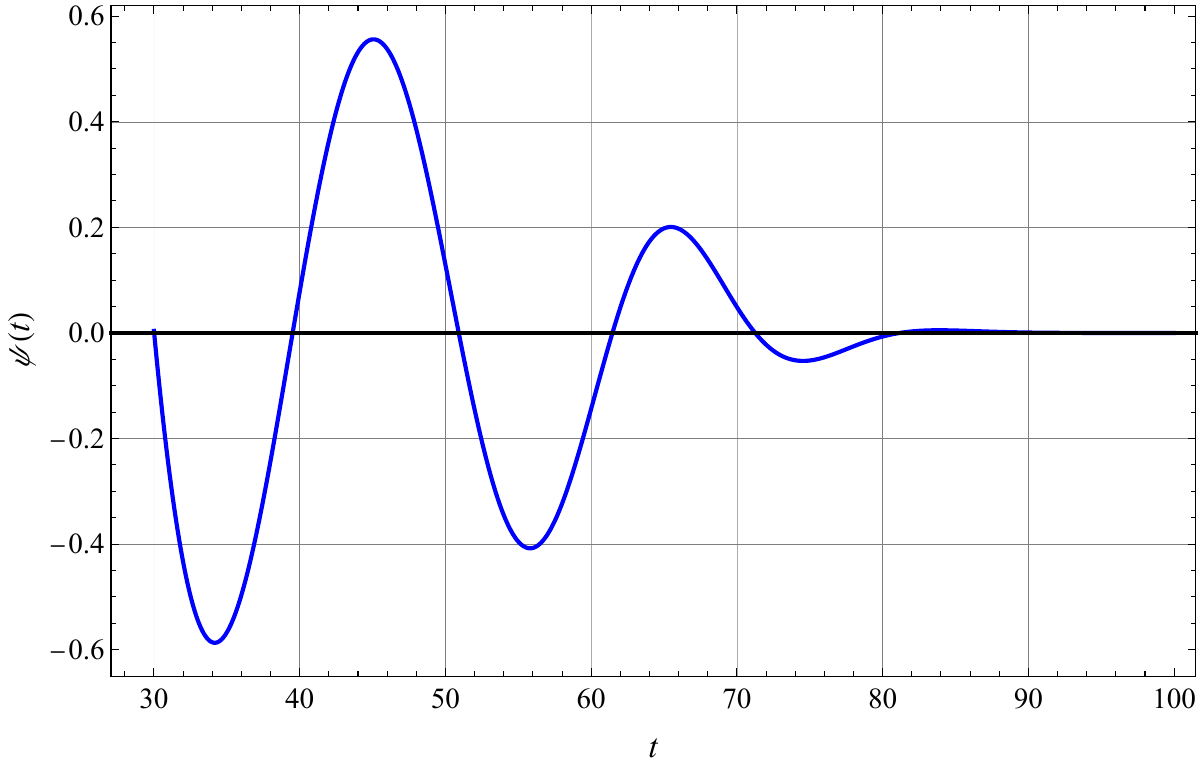}
    }
    \hfill
    \subfigure[]{
        \includegraphics[width=0.47\linewidth]{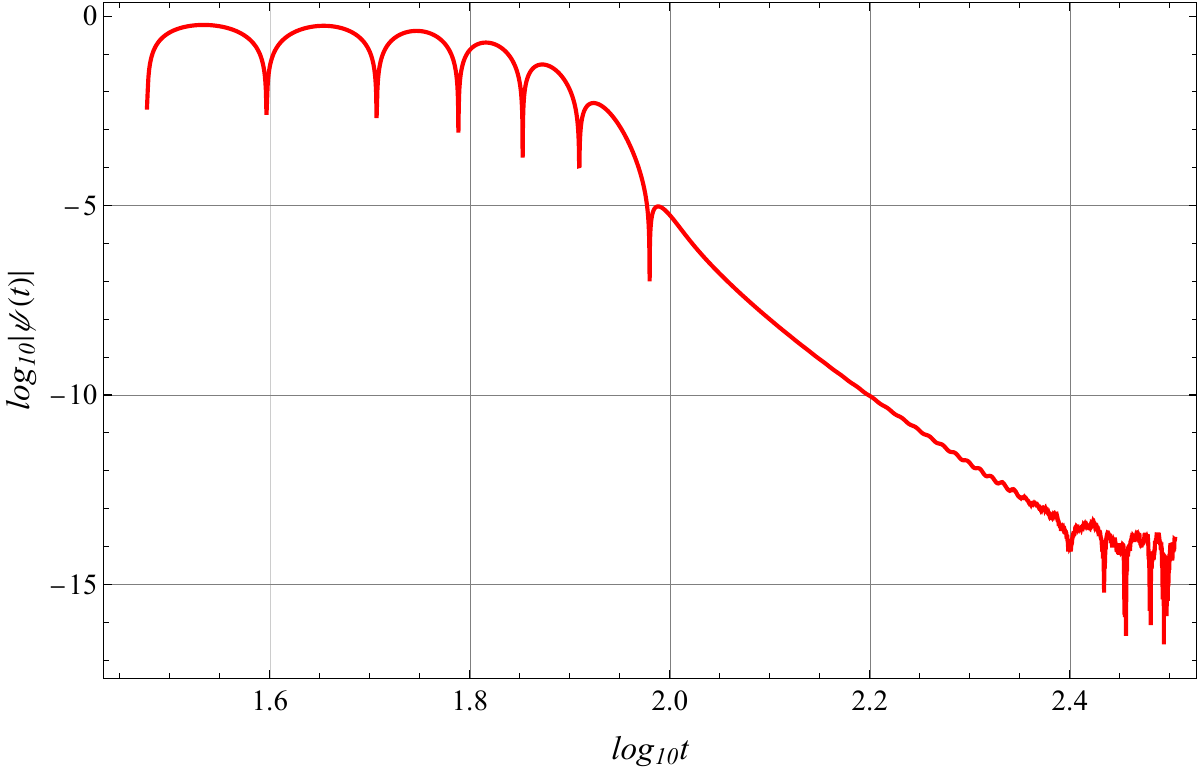}
    }

    \vspace{0.3cm}

    \subfigure[]{
        \includegraphics[width=0.47\linewidth]{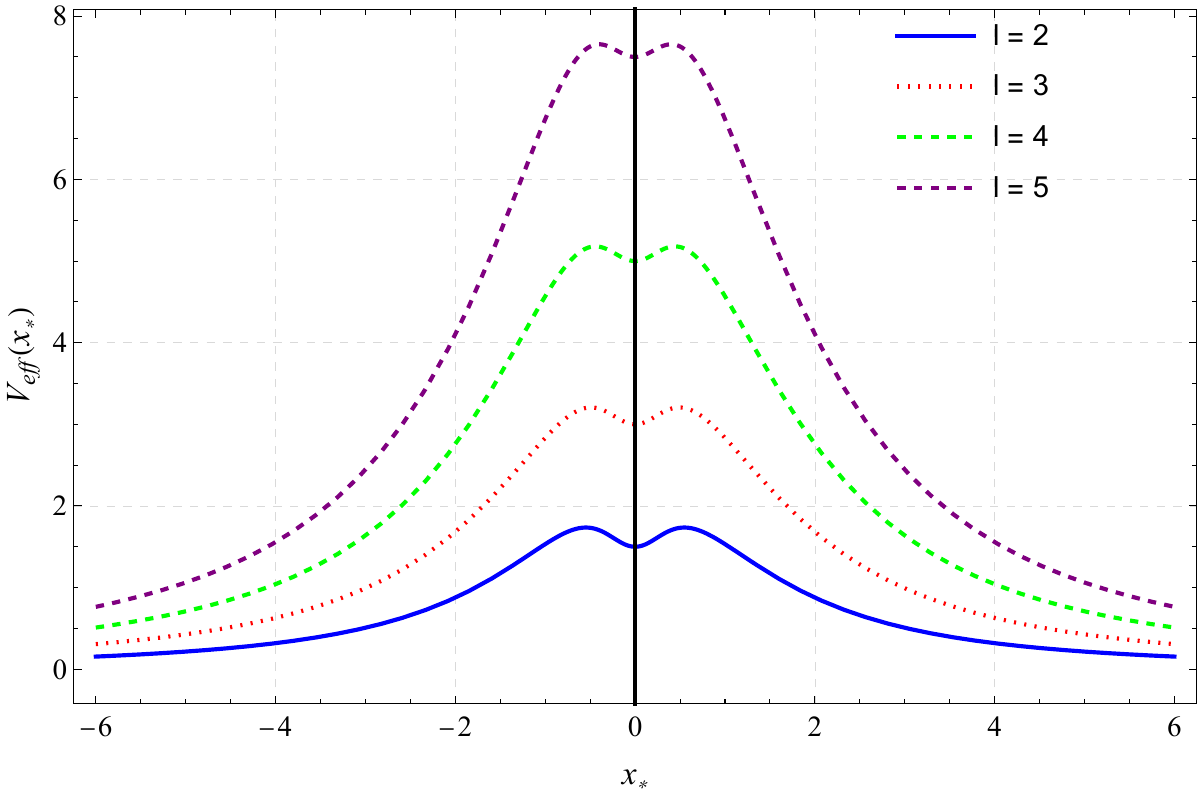}
    }

    \caption{In (a) we have the effective potential of the first model for $a=1,m=2$,
in (b), we have the effective potential of the second model for $d =1 ,b =2 ,c_3=3$,
in (c), we have the test scalar field $\psi$ as a function of time, for $d = 3,m = 1,b = 5,l = 5,\sigma=6,v_c=40$,
in (d) we have the logarithm of the test scalar field $\psi$ as a function of the logarithm of time, for $d = 3,m = 1,b = 5,l = 5,\sigma=6,v_c=40$, and in (e) we have the effective potential of the third model for $d =0.8 ,m = 1 ,b =2$.}
    \label{fig-Veff}
\end{figure}

In general, the three potentials possess two maxima. Therefore, we shall treat the first two cases, namely the first two potentials with $m=2$, which have a single maximum, using the sixth-order WKB approximation, and the last potential, for $m=1$, which possesses two maxima, using the time-domain evolution method. This will provide an indication of stability (or instability). The effective potential plots are shown in Fig. \ref{fig-Veff}.

We shall consider only these cases, since a complete analysis would be far more detailed and would constitute an independent work devoted exclusively to this topic. A more rigorous analysis would require studying all spins, namely spin zero, one, and two, while the two remaining spins not considered here should include contributions from the interaction with the NED. See \cite{Bronnikov2022,Toshmatov2019}, for instance.
 
Applying the sixth-order WKB approximation to the effective potentials \eqref{eqVeff1} and \eqref{eqVeff2} of the test scalar field from equation \eqref{eqSchrodinger}, we obtain the following list of values shown in tables \ref{tab1} and \ref{tab2}, respectively. We observe that all values of the imaginary part of the frequency are negative, indicating linear stability within the sixth-order WKB approximation.
\begin{table}[htbp]
\centering
\caption{Quasinormal modes obtained by the sixth-order WKB method for model I, with $m=2,a=1,n=2$.}

\begin{tabular}{ccccc}
\hline
$\ell$ & $x_{\max}$ & $V_{\mathrm{eff}}(x_{\max})$ & $\mathrm{Re}(\omega)$ & $\mathrm{Im}(\omega)$ \\
\hline
2 & 0 & 7 & 1.9471676560 & -2.7515290330 \\
3 & 0 & 13 & 3.1003348190 & -2.6249354900 \\
4 & 0 & 21 & 4.1898830580 & -2.5733222110 \\
5 & 0 & 31 & 5.2471426760 & -2.5481206660 \\
6 & 0 & 43 & 6.2866120820 & -2.5340551310 \\
7 & 0 & 57 & 7.3154152440 & -2.5254026910 \\
8 & 0 & 73 & 8.3373536590 & -2.5196920880 \\
9 & 0 & 91 & 9.3546204130 & -2.5157202460 \\
10 & 0 & 111 & 10.3685653800 & -2.5128437980 \\
\hline
\end{tabular}\label{tab1}
\end{table}
\begin{table}[htbp]
\centering
\caption{Quasinormal modes obtained by the sixth-order WKB method for model II, with
 $d=1,b=2,n=2,c_3=3$.}

\begin{tabular}{ccccc}
\hline
$\ell$ & $x_{\max}$ & $V_{\mathrm{eff}}(x_{\max})$ & $\mathrm{Re}(\omega)$ & $\mathrm{Im}(\omega)$ \\
\hline
2 & 0 & 2.1371383840 & 10.5135005600 & -2.2469773730 \\
3 & 0 & 3.7187212130 & 2.6692026240 & -4.1272259740 \\
4 & 0 & 5.8274983180 & 1.0627151850 & -4.7642074710 \\
5 & 0 & 8.4634696990 & 1.0714513590 & -3.7334519780 \\
6 & 0 & 11.6266353600 & 1.6960132510 & -2.7202573570 \\
7 & 0 & 15.3169952900 & 2.6006365680 & -2.1801084490 \\
8 & 0 & 19.5345495000 & 3.4232074390 & -1.9882761890 \\
9 & 0 & 24.2792979900 & 4.1313354260 & -1.9198480500 \\
10 & 0 & 29.5512407500 & 4.7711028370 & -1.8923088940 \\
\hline
\end{tabular}\label{tab2}
\end{table}

Let us only check the eikonal limit ($l\gg 1$) of the imaginary part \cite{Cardoso2009}
\begin{align}
&\Omega_c = \frac{1}{\Sigma(x_c)},\,\lambda_c = \sqrt{\frac{\Sigma''(x_c)}{\Sigma(x_c)}},\,\Sigma'(x_c) = 0,\\
&\omega_{QNM} \simeq 
\left(l+\frac12\right)\Omega_c
-i\left(n+\frac12\right)|\lambda_c|.
\end{align}
The conclusion is that the imaginary part of the quasinormal mode frequency is always negative in these cases. Therefore, we have an indication of linear stability.

Now, for the third case, we shall use the time-domain evolution method, equations \eqref{TD1}-\eqref{TD2}. Taking the effective potential \eqref{eqVeff3}, we see that there are two maxima, as shown in Fig. \ref{fig-Veff}. Thus, these two barriers act as cavities for a wave emitted in the direction of the WH. The wave encounters the first barrier in our universe, where part of it is reflected and another part is transmitted. It then passes through the throat and emerges in the other universe, with $x<0$. The wave then encounters the second barrier, where again part of it is reflected and part is transmitted. The reflected component crosses the throat in the opposite direction and encounters the first barrier in our universe for the second time. Hence, part of the wave is reflected and part transmitted. This process continues until all the energy is dissipated. This phenomenon is known as an echo.

We graphically represent the test scalar field $\psi(t)$ and the logarithmic scale, with $x_{*}=30$, Fig. \ref{fig-Veff}. We can see that there are no signs of echoes, and that the logarithmic scale indicates stability. We implemented the code in the Mathematica software, where the numerical precision is of the order of $10^{-15}$. Therefore, in the logarithmic-scale plot, $-15$ actually marks the numerical zero of $\psi(t)$. We conclude that the third model, for the parameter values studied, indicates linear scalar stability. On the other hand, the Prony estimation method yields $\omega \simeq 21.63270 - 2.64194\,i$, indicating linear scalar stability, since $\mathrm{Im}(\omega)<0$.

\section{Conclusion}\label{sec7}

In this work, we start from a general action that describes the k-essence scalar field theory, assuming a power form and coupled to the geometric sector. This theory was used to investigate black-bounce solutions \cite{CDJM1} and generalizations for different scalar field configurations \cite{CDJM2}. We extended this framework by introducing NED \cite{Rodrigues:2023vtm,CDJM3}, where we explored new magnetically charged black-bounce solutions for scalar field powers that differ from the canonical case $n=1$. Specifically, we constructed magnetic solutions such that the throat of the wormhole coincides with the magnetic charge $a=q_m$ and subsequently also applied bumblebee gravity \cite{CDJM4}. Based on the equations of motion, Eqs. (\ref{14}–\ref{17}), it is observed that the electromagnetic function $L_f$, appearing in Eq. (\ref{15}), does not depend on the specific form of the scalar field, but solely on the metric functions $A(x)$ and $\Sigma(x)$. Therefore, it follows that this function coincides with that obtained for a canonical scalar field, being determined exclusively by the chosen model.

For all models investigated in this work, we analytically derived all the functions of interest involved, and for simplicity in this process of solving the equations of motion, and considering that the energy conditions are preserved independently of the power of the field of k-essence \cite{CDJM3,CDJM4}, we fixed $n=1/2$. In general, we can verify that for both electrically and magnetically charged systems, the electromagnetic functions $L_f(x)$ and $L(x)$ do not change with variations in the power of the k-essence field. In fact, these functions are the same as those obtained in the canonical case and investigated in \cite{INTRO19,Rodrigues:2023vtm}. This behavior implies that the modifications in the scalar field due to k-essence are counterbalanced by the potential, keeping $L_f(x)$ and $L(x)$ unchanged.

We initially analyzed the GEB wormhole model (\ref{sec3}) and then obtained the expressions for the phantom scalar field, the potential, and the electromagnetic functions for the magnetically and electrically charged system, for the k-essence field power configuration $n=1/2$. We also analyzed the behavior of these quantities by varying the parameter $m\geq{2}$. We can observe in Fig. (\ref{FIG1MOD1}) that all the quantities under analysis increase in amplitude in the asymptotic $x\to{\pm\infty}$ as the values of the parameter $m$ increase. On the other hand, when we look at the potential, we see that it has a potential barrier-type behavior, which may indicate the possibility of instability when subjected to linear perturbations.

With respect to the second model (\ref{sec4}), which likewise describes the spacetime of a wormhole whose throat can be tuned by calibrating the parameters that define the area function, we obtained the expressions for the phantom scalar field, the potential, and the electromagnetic functions for the two scenarios analyzed. In Fig. (\ref{FIG1MOD2}) we have the representation of these quantities mentioned above as a function of the radial distance. We varied the parameters and we can emphasize, in relation to the potential of this model, that depending on the adjustment between the choice of these parameters $b,d,c_{3}$ we can have configurations in which the potential has two negative minimum points, which may suggest that the regime has the possibility of stability when subjected to perturbations.

Finally, the third model also describes a wormhole. In general, we analytically derived the expressions for the phantom scalar field, responsible for supporting the violation of the energy conditions required to sustain the wormhole structure, as well as the associated potential and the corresponding electromagnetic functions for the two systems investigated.

From the inequality associated with $\mathrm{NEC}_{1}^{\phi}$, we find that the NEC is always violated for the generalized Ellis--Bronnikov model, whereas for the other models it is possible to satisfy this inequality only in certain regions, even with modifications of the model parameters,. Therefore, in a global sense, the NEC is violated in all models and, consequently, since the inequality associated with $\mathrm{NEC}_{1}^{\phi}$ also appears in the remaining energy conditions, all the standard energy conditions are violated.

We have also investigated the linear scalar stability of the wormhole geometries through QNM and time--domain evolution of a test scalar field. For the cases in which the effective potential possesses a single maximum, namely models I with $m=2$ and II, the sixth--order WKB approximation was employed. In all analysed configurations, the imaginary part of the quasinormal frequencies remained negative, indicating damped oscillations and, consequently, linear scalar stability within the adopted approximation. This behaviour is further supported in the eikonal limit, where the imaginary part of the frequencies also remains negative.
For the third model, whose effective potential exhibits two maxima, the WKB approach is no longer directly applicable. In this case, we performed a time--domain analysis using characteristic integration in double--null coordinates. The obtained wave profiles display a decaying behavior without any indication of growing modes, while the logarithmic evolution confirms the stability of the perturbations. In addition, the Prony extraction method yields frequencies with negative imaginary part, reinforcing the indication of linear scalar stability for the parameter range investigated.
It is important to emphasise that the present analysis is restricted to scalar perturbations and selected regions of the parameter space. A complete perturbative study would require the investigation of electromagnetic and gravitational perturbations, particularly because the nonlinear electrodynamics sector may contribute nontrivially to higher--spin perturbations. Therefore, the results presented here should be interpreted as indications of linear stability rather than a fully exhaustive proof.

In contrast to previous studies in the literature that focus on black hole solutions \cite{NOVA1,NOVA2}, the present work is devoted to establishing existence and stability criteria for a class of wormhole geometries. Moreover, our approach emphasizes the construction of new solutions and the identification of their associated matter sources in astrophysical contexts. By comparison, other investigations, such as those reported in Refs. \cite{INTRO42,NOVA4}, are primarily aimed at examining the implications of k-essence theory within cosmological scenarios. Our work also reveals that the matter sources supporting wormholes discussed in the literature are not unique. Models that are commonly represented by a standard but phantom scalar field \cite{EB3} can also be generated by alternative frameworks, such as $k$-essence models. Furthermore, our work goes a step further by investigating the stability of these configurations through QNM and the time-domain evolution method.

It is well known that the presence of nonlinear electrodynamics can alter the photon trajectories, thereby modifying the shadow and the optical appearance of the solutions \cite{INTRO12}. Accordingly, one of our planned future works is to investigate how the sources found in this paper can change the optical appearance of these solutions. Another important point is that the stability of the solutions also depends on the type of source considered, so both stable and unstable solutions may occur. In this way, as future work, we intend to perform a more detailed stability analysis of these models following the procedure adopted in \cite{Franzin:2023slm}, by perturbing all the fields involved in the solution in order to understand how the choice of matter sources can affect the stability of these models. \cite{INTRO43,Bronnikov:2013coa,Nandi:2016ccg}.

The scalar field potential shows parameter regions with possible stability (the potential $V(\phi)$ has a minimum, $m_{eff}^2>0$), though a detailed, model-dependent analysis is still required. The EB wormhole is generally unstable under geometrical perturbations \cite{gonzalez2009instability,bronnikov2011stability,bronnikov2012instabilities,bronnikov2004conformal,bronnikov2007no}, but alternative approaches indicate possible stability \cite{Nandi:2016ccg}, which may also apply to our models. Wormholes supported by phantom matter can be matched to stable exterior solutions, leading to stable configurations \cite{lobo2005stability}. In NED, the form of the Lagrangian $L(f)$ can enhance or suppress instabilities, allowing for stable solutions \cite{Bronnikov2022}. We believe that our cases may also exhibit stability for an appropriate choice of parameters. Our solutions can be matched to an external black hole solution in such a way that our structures represent an interior solution, resembling a regular anisotropic expanding universe, commonly referred to as a black-universe configuration of the Kantowski–Sachs cosmological type \cite{bronnikov2007regular}.

\begin{acknowledgments}
We thank CNPq, CAPES, FAPES, and FUNCAP for financial support.
\end{acknowledgments}

\clearpage

\nocite{*}
		
\end{document}